\documentclass[
preprint,superscriptaddress,
 amsmath,amssymb,
 aps,
prb,
floatfix,
]{revtex4-2}

\usepackage{graphicx}% Include figure files
\usepackage{dcolumn}% Align table columns on decimal point
\usepackage{bm}% bold math
\usepackage{braket}
\usepackage{relsize}
\usepackage{xcolor}
\usepackage{hyperref}% add hypertext capabilities
\hypersetup{
    colorlinks=true,
    linkcolor=blue,
    citecolor=blue,      
    urlcolor=blue
    }

\begin{document}

\title{Semiclassical Electron and Phonon Transport from First Principles: Application to Layered Thermoelectrics}

\author{Anderson S. Chaves}
\affiliation{School of Engineering and Applied Sciences, Harvard University, Cambridge, Massachusetts 02138, United States}

\author{Michele Pizzochero}
\affiliation{School of Engineering and Applied Sciences, Harvard University, Cambridge, Massachusetts 02138, United States}

\author{Daniel T. Larson}
\affiliation{Department of Physics, Harvard University, Cambridge, Massachusetts 02138, United States}

\author{Alex Antonelli}
\affiliation{Gleb Wataghin Institute of Physics and Centre for Computational Engineering \& Sciences, University of Campinas, UNICAMP, 13083-859 Campinas, S\~{a}o Paulo, Brazil}

\author{Efthimios Kaxiras}
\affiliation{School of Engineering and Applied Sciences, Harvard University, Cambridge, Massachusetts 02138, United States}
\affiliation{Department of Physics, Harvard University, Cambridge, Massachusetts 02138, United States}

\date{\today}

\begin{abstract}
Thermoelectrics are a promising class of materials for renewable energy owing to their capability to generate electricity from waste heat, with their performance being governed by a competition between charge and thermal transport. 
A detailed understanding of energy transport at the nanoscale is thus of paramount importance for developing efficient thermoelectrics.  
Here, we provide a comprehensive overview of the methodologies adopted for the computational design and optimization of thermoelectric materials from first-principles calculations. First, we introduce density-functional theory, the fundamental tool to describe the electronic and vibrational properties of solids. 
Next, we review charge and thermal transport in the semiclassical framework of the  Boltzmann transport equation, with a particular emphasis on the various scattering mechanisms between phonons, electrons, and impurities. 
Finally, we illustrate how these approaches can be deployed in determining the figure of merit of tin and germanium selenides, an emerging family of layered thermoelectrics that exhibits a promising figure of merit. Overall, this review article offers practical guidelines to achieve an accurate assessment of the thermoelectric properties of materials by means of computer simulations. 
\end{abstract}

\maketitle
\newpage

\tableofcontents

\section{Introduction}

Navigating the global challenges for sustainable energy production, distribution, and use requires continued exploration of new and tailor-made materials to increase efficiency.
Detailed knowledge and understanding of electrical and thermal transport properties allows for their optimization, hastening technological development. 
Since thermoelectric devices can convert waste heat to electricity without any mechanical parts, developing improved thermoelectric materials is a promising approach for energy efficiency and reliability \cite{yang2018high}.

This paper provides an overview of the first-principles tools that can be used to calculate the transport properties of layered crystal structures.
Starting from the density functional theory (DFT) approach to calculating the ground state of the many-body quantum mechanical electron-nuclear Hamiltonian, we describe how various scattering rates, including electron-phonon, phonon-phonon, polar, and defect scattering, can be determined and used as input to the semiclassical Boltzmann Transport Equation (BTE) in order to calculate transport properties such as electrical and thermal conductivity.

\begin{figure}[t]
\includegraphics[width=\columnwidth]{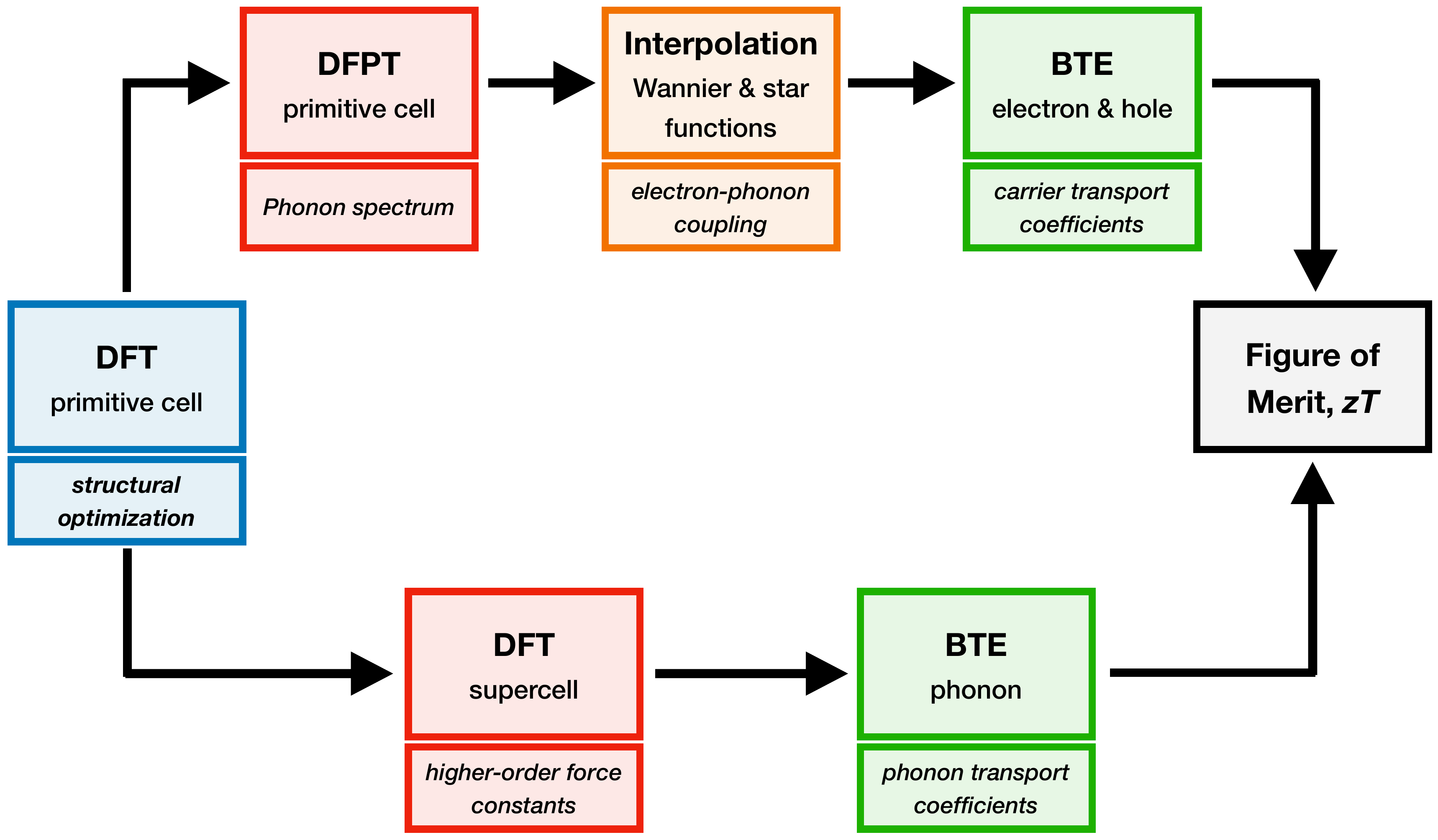}
\caption{\label{fig:workflow} Computational workflow for the calculation of the thermoelectric figure of merit, $zT$, using first-principles methods based on density functional theory (DFT).}
\end{figure}

To demonstrate the power of this framework we focus on the thermoelectric figure of merit, $zT$. Not only is the study and optimization of thermoelectric materials a vibrant and timely research direction due to the importance of thermoelectrics for energy recovery from waste heat, the determination of $zT = \sigma S^2 T / (\kappa_\mathrm{carr} + \kappa_\mathrm{latt})$, requires knowledge of several distinct transport properties that are amenable to first-principles calculations, namely the electrical conductivity $\sigma$, Seebeck coefficient $S$, carrier (electron or hole) thermal conductivity $\kappa_\mathrm{carr}$, and the lattice thermal conductivity $\kappa_\mathrm{latt}$, all for a given temperature $T$.
A high-level workflow for the calculation of $zT$ is shown in Fig.~\ref{fig:workflow}. Our aim in the following is to give some more detailed explanations of both the conceptual and practical aspects of each of the steps in the calculation. Thermoelectric research is a vast field, both on the theoretical and experimental sides, and a thorough review of the subject is beyond the scope of the present work. Other recent reviews of thermoelectric materials\cite{yang2018high,hasan2020inorganic}, applications\cite{zoui2020review}, and computational approaches\cite{gutierrez2020review} can provide a broader context. Here we focus on the calculation of $zT$ for a specific class of promising layered materials as an example for how the previous techniques can be put into practice.

The rest of the article is organized as follows. In Sect.~\ref{sec:theory} we concisely introduce the basic principles of DFT, the computational tool that underlies these first-principles calculations, and also density functional perturbation theory (DFPT). In Sect.~\ref{sec:bte} we present the Boltzmann Transport Equation, while Sect.~\ref{sec:scattering-mech} describes in detail several important scattering mechanisms. Sect.~\ref{sec:numerics} gives details on the post-processing of the DFT results, and Sect.~\ref{sec:application} describes the application of these computational tools in the prediction of $zT$ for SnSe and GeSe, including new results for the lattice thermal conductivity. Finally, we conclude in Sect.~\ref{sec:conclusion}.

\section{Theoretical framework}\label{sec:theory}

\subsection{Kohn-Sham (KS) Density Functional Theory} 

Owing to the relative simplicity, reliability, reasonable computational effort, and extensive implementation in widely available software packages \cite{Giustino2014}, DFT has rapidly emerged as the method of choice to find approximate solutions to the many-body problem and determine the properties of molecules and materials at the quantum-mechanical level \cite{Szabo1996}.  DFT is conceptually rooted in the two theorems introduced by Hohenberg and Kohn \cite{Hohenberg1964}, and its practical deployment follows the KS Hamiltonian \cite{Kohn1964}, 

\begin{equation}
\label{HKS}
\hat{H}^{KS} = -\frac{1}{2} \nabla^2 + V(\mathbf{r}) + V_\textsubscript{H}(\mathbf{r}) + V^{xc}(\mathbf{r}),
\end{equation}
where $V(\mathbf{r})$ is the external potential, $V_\textsubscript{H}(\mathbf{r})$ the Hartree potential, and $V^{xc}(\mathbf{r})$ the exchange-correlation potential. This Hamiltonian leads to a set of $n$ Schr\"odinger-like equations for the single-particle Kohn-Sham orbitals $\Psi (\mathbf{r})$ and accompanying energies $\epsilon$,
\begin{equation}
\hat{H}^{KS} \Psi_i (\mathbf{r}) = \epsilon_i \Psi_i(\mathbf{r}).
\end{equation}
The solution of the Kohn-Sham equations proceeds in self-consistent fashion by  (i) starting with an educated guess of the electron density $n_k(\mathbf{r})$ to (ii) construct the Kohn-Sham Hamiltonian $\hat{H}^{KS}$, (iii) finding upon diagonalization the corresponding eigenvalues $\epsilon_i$ and eigenvectors, which are subsequently used to (iv) obtain the new density $n_{k+1}(\mathbf{r})$, until the difference between $n_{k+1}(\mathbf{r})$ and $n_k(\mathbf{r})$ does not exceed a given numerical tolerance. 

A number of approximations to the exchange-correlation potential appearing in Equation \ref{HKS} has been devised \cite{Sholl2009,ParrYang1994}, most notably (i) the Local Density Approximation (LDA) \cite{LDA}, in which the electron density is assumed to be locally the same as a spatially uniform electron gas with the same density, (ii) the Generalized Gradient Approximation (GGA) \cite{PBE, PW91, PBEsol, revPBE}  and Meta-Generalized Gradient Approximation (meta-GGA) \cite{RTPSS, SCAN}, and Laplacian of the electron density or the kinetic energy density, are taken into account to increase the accuracy over LDA, and (iii) hybrid functionals \cite{PBE0, HSE03, HSE06}, in which a fraction of the orbital-dependent, exact Fock exchange is included in the otherwise bare GGA.

Although the exchange interactions between electrons can be determined exactly (cf.\ Hartree-Fock equations), correlation interactions are invariably approximated, a trait that is common to all the exchange-correlation functionals listed above. It is therefore not surprising that density-functional theory provides an inaccurate account of those physical situations where correlation effects dominate, for example, van der Waals interactions or localized electronic states. In addition, being DFT commonly considered a ground state theory (cf.\ Hoehnberg-Kohn theorems), predictions concerning excited states and band gaps are often quantitatively unreliable, see references \cite{gorling1996density,cohen2008insights} for through discussions.

\begin{table*}[t]
\begin{tabular}{@{}|c|c|c|c|@{}}
\hline
\textbf{Software} & \textbf{License} & \textbf{Reference} & \textbf{Webpage} \\ 
\hline
ABINIT & Free, GPL & \cite{Gonze2009} & \url{www.abinit.org} \\ 
CASTEP & Academic, Commercial & \cite{Clark2005} & \url{www.castep.org} \\ 
CP2K & Free, GPL & \cite{CP2K} & \url{www.cp2k.org} \\ 
GPAW & Free, GPL & \cite{Enkovaara2010} &   \url{wiki.fysik.dtu.dk/gpaw}  \\
ONETEP & Academic, Commercial & \cite{Prentice2020} & \url{www.onetep.org}  \\ 
Quantum ESPRESSO & Free, GPL & \cite{Giannozzi_2009} & \url{www.quantum-espresso.org}  \\ 
VASP & Academic, Commercial & \cite{kresse1996efficient,kresse1996efficiency} & \url{www.vasp.at} \\
\hline
\end{tabular}
\caption{A list of representative software packages implementing plane-wave DFT for solid-state systems. \label{DFT-TABLE}}
\end{table*}

%When core electrons are described through pseudopotentials, one is left with valence electrons, the wavefunctions $\phi(\mathbf{r})$ of which are expanded on a finite basis. 

When periodic boundary conditions apply, like in the case of crystalline materials, one can take advantage of the Bloch theorem, which allows the single-particle wavefunctions to be labeled by a band index $n$ and the crystal momentum $\mathbf{k}$ and written as the product of a plane wave and a function that is periodic in the crystal lattice:
\begin{equation}
\phi_{n,\mathbf{k}}(\mathbf{r}) = \psi_{n,\mathbf{k}}({\mathbf{r}})  e^{i\mathbf{k} \cdot \mathbf{r}},\qquad \psi_{n,\mathbf{k}}(\mathbf{r}+\mathbf{R}) = \psi_{n,\mathbf{k}}(\mathbf{r})\, .
\label{PW}
\end{equation}
 Because of this periodicity, $\psi_{n,\mathbf{k}}(\mathbf{r})$ can be expanded in a Fourier series including only reciprocal lattice vectors, $\mathbf{G}$:
\begin{equation}
    \psi_{n,\mathbf{k}}(\mathbf{r}) = \sum_\mathbf{G} e^{i\mathbf{G}\cdot\mathbf{r}} \psi_{n,\mathbf{k}}(\mathbf{G})\,.
\end{equation}
Thus, plane waves are the most natural basis set for periodic systems. According to Eq.~(\ref{PW}), however, the evaluation of the solution at a single point requires the summation over an infinite number of $\mathbf{G}$ vectors. In practice, this expansion is truncated to an energy cutoff, $E\textsubscript{cut}$, implying that only solutions  of kinetic energy
\begin{equation}
\frac{h^2}{2m_e} | \mathbf{k} + \mathbf{G} | ^2 < E\textsubscript{cut}
\end{equation}
are evaluated. Delocalized plane waves further offer the advantage of completeness, and the convergence of the calculated properties can be systematically improved by increasing the energy cutoff \footnote{Besides plane waves, localized basis sets consisting of atomic-like orbitals (e.g., Gaussian- or Slater-type functions) have found a widespread use, in particular in the computational chemistry community. Contrary to plane waves, fewer basis functions are often needed to achieve a reasonable accuracy, hence significantly decreasing the computational effort. However, localized basis sets are controlled by many parameters in addition to the energy cutoff, in a way that no systematic convergence can be attained.}. A list of representative software packages implementing plane-wave density-functional theory for solid-state systems is given in Table \ref{DFT-TABLE}.

Although extended Bloch orbitals are the most natural choice to describe the electronic states of a periodic system, an alternative approach based on the Wannier representation is convenient for an array of applications in electronic structure \cite{marzari1997maximally, Marzari2012}. Contrary to Bloch orbitals, Wannier functions provide a real-space representation of localized orbitals which are specified by a lattice vector $R$ and a band index $n$.  The general transformation from Bloch orbitals into Wannier functions is
\begin{equation}
\label{eqn:mlwf}
\ket{W_{\mathbf{R},n}} = \frac{V}{(2\pi)^2} \int_\textnormal{BZ} d\mathbf{k} \sum_{m=1}^{N} U_{m,n}^\mathbf{k} \psi_{m,\mathbf{k}} e^{{-i\mathbf{k}\cdot \mathbf{R}}},
\end{equation}
where $V$ is the volume of the primitive cell and $U_{m,n}^\mathbf{k}$ is a gauge-fixing, $N$-dimensional unitary matrix. Additional freedom in choice of $U_{m,n}^\mathbf{k}$ means that the Wannier functions are not uniquely determined. A popular choice for the gauge is determined by minimizing the spread functional, $\Omega$:
\begin{equation}
\frac{\partial \Omega [U]}{\partial U} = 0, 
\end{equation}
where
\begin{equation}
\Omega = \sum_n \left[ \bra{W_{0,n}}r^2\ket{W_{0,N}} - \bra{W_{0,n}}r\ket{W_{0,n}}^2 \right]. 
\end{equation}
 Wannier functions obtained following this procedure are referred to as Maximally Localized Wannier Functions (MLWF). Wannierization of DFT calculations is typically performed with the Wannier90 package \cite{Mostofi2008, Mostofi2015}.

\subsection{Lattice dynamics from density functional perturbation theory}
\label{sec:dfpt}

Within the Born-Oppenheimer (BO) approximation, 
the lattice-dynamical properties of a system 
can be determined by obtaining the eigenvalues, $E$, 
and the eigenvectors, $\Phi$, of the following Schr\"{o}dinger
equation \cite{born1966dynamical}

\begin{equation}
\label{Ion}
\left( \sum_I \frac{\hbar^2}{2M_I}\frac{\partial^2}{\partial {\bf{R}}_I^2} + \hat{H}_{BO}({\bf{R}})\right) \Phi({\bf{R}}) = E({\bf{R}})\Phi({\bf{R}})~.
\end{equation}

Here, the index $I$ labels each ion in the system, 
while ${\bf{R}}_I$ and $M_I$ correspond to its coordinates and masses, respectively. 
For conciseness, $\mathbf{R}$ correspond to all ionic coordinates and  
$\hat{H}_{BO}$ is the BO electronic Hamiltonian that depends parametrically upon $\mathbf{R}$
\begin{equation}
\label{HBO}
\hat{H}_{BO}({\bf{R}}) = \frac{-\hbar^2}{2m}\sum_i \frac{\partial^2}{\partial{\bf{r}}_i^2} + \frac{e^2}{2}\sum_{i \neq j} \frac{1}{|{\bf{r}}_i - {\bf{r}}_j|} - \sum_{iI} \frac{Z_Ie^2}{|{\bf{r}}_i - {\bf{R}}_I|} + \frac{e^2}{2} \sum_{I \neq J} \frac{Z_IZ_J}{|{\bf{R}}_I - {\bf{R}}_J|}~.
\end{equation}
Interatomic forces are computed from the derivative 
of the ground-state energy with respect to the ion positions, and
can be calculated by using the Hellmann-Feynman theorem \cite{hellmann1937einfuhrung,feynman1939forces}:
\begin{eqnarray}
\label{HF}
{\bf{F}}_I = -\frac{\partial E({\bf{R}})}{\partial {\bf{R}}_I} &=& 
-\braket{\Phi({\bf{R}})|\frac{\partial(\hat{H}_{BO})}{\partial {\bf{R}}_I}|\Phi({\bf{R}})} \\
\label{forces}
 &=& -\int n_{{\bf{R}}}({\bf{r}}) \frac{\partial V_{{\bf{R}}}({\bf{r}})}{\partial {{\bf{R}}}_I}d{\bf{r}} - \frac{\partial E_N({{\bf{R}}})}{\partial {{\bf{R}}}_I}~,
\end{eqnarray}
where ${\bf{R}}_I = {\bf{R}}_p + \xi_{\kappa\alpha p}$, with 
${\bf{R}}_p$ one of the direct lattice vectors of a 
Born-von K{\'a}rm{\'a}n supercell composed of $N_p$ unit cells of the crystal, and $\xi_{\kappa\alpha p}$
is the position vector with Cartesian coordinate $\alpha$ of the nucleus $\kappa$ in the unit cell labeled by $p$. 
In Eq.~(\ref{forces}) we have 
\begin{equation}
\label{pot}
V_{{\bf{R}}}({\bf{r}}) = \sum_{iI} \frac{Z_Ie^2}{|{\bf{r}}_i - {\bf{R}}_I|}~,
\end{equation}
which is the external potential, and $n_{{\bf{R}}}({\bf{r}})$ is the ground-state electron charge density for a given geometry. $E_N({{\bf{R}}})$ is the electrostatic interaction between nuclei
 \begin{equation}
\label{EN}
E_N({{\bf{R}}}) = \frac{e^2}{2}\sum_{I \neq J} \frac{Z_I Z_J}{|{\bf{R}}_I - {\bf{R}}_J|}~.
\end{equation}
Equation~(\ref{forces}) assumes the external potential acting on the electrons is a differentiable function of
the crystal coordinates.
In particular, the equilibrium geometry is reached when forces acting on individual ion vanish, that is, 
${\bf{F}}_I \equiv \frac{\partial E({\bf{R}})}{\partial {\bf{R}}_I} = 0$. The vibrational properties 
of a system can be determined by the interatomic force constants, 
which correspond to the second derivatives of $E({\bf{R}})$ in relation to ion displacements and 
can be calculated via Hellmann-Feynman theorem as well
\begin{eqnarray}
\label{Hessian}
{\bf{C}}_{\kappa\alpha p, \kappa^{\prime} \alpha^{\prime} p^{\prime}} &=& \frac{\partial^2E}{\partial {\bf{\xi}}_{\kappa\alpha p}\partial {\bf{\xi}}_{\kappa^{\prime} \alpha^{\prime} p^{\prime}}}  \\
{\bf{C}}_{I,J} &=& \frac{\partial^2E({{\bf{R}}})}{\partial {\bf{R}}_I\partial {\bf{R}}_J} =  -\frac{\partial {\bf{F}}_I}{\partial {\bf{R}}_J} \\
&=& \int \frac{\partial n_{{\bf{R}}}({\bf{r}})}{\partial {\bf{R}}_J} \frac{\partial V_{{\bf{R}}}({\bf{r}})}{\partial {\bf{R}}_I} d{\bf{r}} + 
\int n_{{\bf{R}}}({\bf{r}}) \frac{\partial^2 V_{{\bf{R}}}({\bf{r}})}{\partial {\bf{R}}_I \partial {\bf{R}}_J} d{\bf{r}} + 
\frac{\partial^2 E_N ({\bf{R}})}{\partial {\bf{R}}_I \partial {\bf{R}}_J} ~,
\label{eq:hessian3}
\end{eqnarray}
where $\frac{\partial n_{{\bf{R}}}({\bf{r}})}{\partial
{\bf{R}}}$ is the linear response of $n_{{\bf{R}}}({\bf{r}})$
to a distortion of the nuclear geometry, which is a fundamental result derived by De Cicco and
Johnson \cite{decicco1969quantum} and by Pick, Cohen, and Martin \cite{pick1970microscopic}.
This establishes that, within the adiabatic approximation, 
the electrons feel only a static phonon perturbation. 
It is important to note that, due to translational invariance, the interatomic force constants depend on $p$ and $p^{\prime}$
only through the difference ${\bf{R}}_p-{\bf{R}}_{p^{\prime}}$.
The Fourier transform of the interatomic force constants
\begin{equation}
\label{Fouriertransform}
D_{\kappa\alpha,\kappa^{\prime} \alpha^{\prime}}({\bf{q}}) = \frac{1}{\sqrt{M_{\kappa}M_{\kappa^{\prime}}}} \sum_{p} {\bf{C}}_{\kappa\alpha 0, \kappa^{\prime} \alpha^{\prime} p} e^{i{\bf{q}}\cdot{\bf{R}}_p}~,
\end{equation}
is the (hermitian) dynamical matrix \cite{maradudin1968symmetry}, where $M_{\kappa}$ is the mass of the ion $\kappa$. 
Because of the translational invariance, the lattice distortion is monochromatic, meaning that
a phonon perturbation with wavevector $q$ does not induce a force response with wave vector ${\bf{q^{\prime}}} \ne {\bf{q}}$.
The phonon eigenfrequencies of the phonon mode $\nu$ and wave vector ${\bf{q}}$, $\omega_{\nu {\bf{q}}}$, are obtained by 
diagonalizing the dynamical matrix
\begin{equation}
\label{secular}
\sum_{\kappa^{\prime} \alpha^{\prime}} D_{\kappa\alpha,\kappa^{\prime} \alpha^{\prime}}({\bf{q}}) e_{\kappa^{\prime} \alpha^{\prime},\nu}({\bf{q}}) = \omega^2_{\nu {\bf{q}}} e_{\kappa \alpha,\nu}({\bf{q}})~,
\end{equation}
and the eigenvectors $e_{\kappa^{\prime} \alpha^{\prime},\nu}({\bf{q}})$ are the phonon polarizations. 

We now briefly introduce density functional perturbation theory (DFPT), 
which can be used to calculate the interatomic force constants and
vibrational spectra as well as the electron-phonon matrix elements.
As presented in the last section, 
DFT allows for the calculation of the total energy and charge density, which 
can in turn be used to obtain a large number of experimental observables. 
DFPT allows one to obtain quantities that depend on derivatives of the total energy 
and charge density with respect to a small change in the potential.
In particular, DFPT enables the calculation of different properties in condensed matter physics such as vibrational frequencies, elastic constants, dielectric
tensors, Born effective charges, piezoelectric tensors and flexoelectricity \cite{baroni1987elastic,levine1989linear,giannozzi1991ab,de1989piezoelectric,de1991structure,dal1994density,quong1993first,stengel2013flexoelectricity,dreyer2018current,royo2019first}.

First-order perturbative approaches within DFT were proposed independently by 
different groups working in different contexts \cite{stott1980linear,zangwill1980resonant,mahan1980modified,ghosh1982dynamic,zein1984density,baroni1987green}. 
Specifically, the investigation of perturbations in condensed matter physics, 
such as the response of fermionic systems to displacement of atoms within DFT, 
has been pionered by Zein \cite{zein1984density}, Baroni, Giannozzi and Testa \cite{baroni1987green}, 
and Gonze \cite{gonze1992dielectric}. 
A generalization to arbitrary order
was introduced by Gonze and Vigneron \cite{gonze1989density}, on the basis
of the $2n + 1$ theorem of perturbation theory.\cite{hirschfelder1964advances} Such perturbative approaches are based on different techniques comprising 
Green's functions \cite{baroni1987green}, the generalized Sternheimer equation \cite{mahan1980modified,baroni2001phonons}, or the Hylleraas variational scheme \cite{gonze1995adiabatic,gonze1995perturbation}.
Below we will follow the generalized Sternheimer approach. Connections to other techniques have been discussed extensively elsewhere \cite{gonze1995adiabatic}.  

Equation~(\ref{eq:hessian3}) demonstrates that the matrix of force constants
are determined by the electron-density linear response, $\partial n$. Here we will write the KS eigenfunctions $\Psi_{n {\bf{k}}}$ in the Bloch form
\begin{equation}
\label{KSf}
\Psi_{n {\bf{k}}}({\bf{r}}) = N_p^{-1/2} \psi_{n {\bf{k}}}({\bf{r}})e^{i{\bf{k}}\cdot{\bf{r}}}~.
\end{equation}
with lattice periodic part $\psi_{n {\bf{k}}}$.
The density $n$ is given by
\begin{equation}
\label{density1}
n({\bf{r}}) = \sum_{v {\bf{k}}} |\Psi_{v {\bf{k}}}({\bf{r}})|^2
\end{equation}
where the band index $v$ indicates occupied states only. 

Within DFPT, $\partial n$ is induced by the first-order 
variation of the KS potential, $\partial_{\kappa\alpha,{\bf{q}}} v^{KS}({\bf{r}}) e^{i{\bf{q}}\cdot{\bf{r}}}$, 
and can be calculated through the first-order variation of the lattice periodic KS wave 
functions, $\partial \psi_{n{\bf{k}},\bf{q}} e^{i{\bf{q}}\cdot{\bf{r}}}$:
\begin{equation}
\label{density2}
\partial n_{\kappa\alpha,{\bf{q}}}({\bf{r}}) = \frac{2}{N_p}\sum_{v {\bf{k}}}\psi_{v {\bf{k}}}^{\ast} \partial\tilde{\psi}_{v {\bf{k}}, {\bf{q}}}.
\end{equation}
In Eq.~(\ref{density2}), $\partial\tilde{\psi}_{v\mathbf{k},\mathbf{q}} = P^C \partial\psi_{v\mathbf{k},\mathbf{q}}$ is the 
projection of the first-order variation of the lattice periodic KS wave functions onto empty states, where $P^C = (1 - P^V)$ and $P^V = \sum_{v} \ket{\psi_{v {\bf{k}}+{\bf{q}}}} \bra{\psi_{v {\bf{k}}+{\bf{q}}}}$ is the projector onto filled valence states. 

The first-order expansion of the KS equations gives the Sternheimer equation \cite{gonze1995adiabatic}
\begin{equation}
\label{Sternheimer}
\left(\hat{H}^{KS}_{{\bf{k}}+{\bf{q}}} - \epsilon_{v {\bf{k}}}\right)\partial\psi_{v {\bf{k}}, {\bf{q}}} = -\left(\partial_{\kappa \alpha {\bf{q}}} v^{KS} - \partial \epsilon_{v {\bf{k}}}\right)\psi_{v{\bf{k}}}~,
\end{equation}
in which the unperturbed wave functions were written by explicitly indicating the wavevector ${\bf{k}}$
and band index $v$, while the perturbed wavefunctions are projected onto the
manifold of wave vectors ${\bf{k}} + {\bf{q}}$. 
$\hat{H}^{KS}_{{\bf{k}}+{\bf{q}}} = e^{-i({\bf{k}}+{\bf{q}})\cdot{\bf{r}}} \hat{H}^{KS} e^{i({\bf{k}}+{\bf{q}})\cdot{\bf{r}}}$ 
where $\hat{H}^{KS}$ is the unperturbed KS Hamiltonian, Eq.~(\ref{HKS}), and $\partial_{\kappa\alpha,{\bf{q}}}v^{KS}$ 
is the first-order variations of the KS potential 
\begin{equation}
\label{DeltaVBar}
\partial_{\kappa \alpha {\bf{q}}} v^{KS}({\bf{r}}) = \partial_{\kappa \alpha {\bf{q}}} V({\bf{r}}) + e^2 \int \frac{\partial_{\kappa \alpha {\bf{q}}} n ({\bf{r'}})}{|{\bf{r}}-{\bf{r'}}|} e^{-i{\bf{q}}\cdot({\bf{r}}-{\bf{r^{\prime}}})} d{\bf{r'}} + \frac{d_{\kappa \alpha {\bf{q}}} V^{xc}}{dn}\bigg|_{n=n({\bf{r}})} \partial_{\kappa \alpha {\bf{q}}} n ({\bf{r}})~. 
\end{equation}
By expressing $\partial n$ in terms of a sum over the whole spectrum of the
unperturbed Hamiltonian, both occupied and empty states, it becomes evident that 
contributions to $\partial n$ coming from only occupied states will
vanish \cite{giannozzi1991ab}.
Only perturbations that couple the occupied-state
manifold with the empty-state manifold will contribute. 
At the same time, the right-hand side of Eq.~(\ref{Sternheimer}) depends only 
on occupied states, while its left-hand side is ill-conditioned because the possibility of null eigenvalues 
of the linear operator.
In order to make Eq.~(\ref{Sternheimer}) nonsingular, 
both sides are projected onto the empty conduction states by the projector $P^C$.
Additionally, by adding $P^V P^C \partial\psi_{v} = 0$ to the left-hand side of Eq.~(\ref{Sternheimer}) we remove any null eigenvalues without changing the equation:
\begin{equation}
\label{Stern2}
\left(\hat{H}^{KS}_{{\bf{k}}+{\bf{q}}} +\alpha P^V_{{\bf{k}}+{\bf{q}}}- \epsilon_{v {\bf{k}}}\right)\partial\tilde{\psi}_{v {\bf{k}}, {\bf{q}}} = -(1-P^V_{{\bf{k}}+{\bf{q}}})\partial_{\kappa\alpha,{\bf{q}}}v^{KS}\psi_{v{\bf{k}}}~.
\end{equation}

In practice, Eqs.~(\ref{density2}),~(\ref{DeltaVBar}) and~(\ref{Stern2}) are solved self-consistently, just like
the KS equations within DFT. They form a generalized linear problem, since the
KS perturbation, $\partial v^{KS}$, depends linearly on the 
electron-density linear response, $\partial n$, which in turn is 
a linear functional of the variation of the KS wavefunctions, $\partial \psi$. 
In practice, the initial KS perturbation is set to be equal to the external potential, Eq.~(\ref{pot}). 
By solving the Sternheimer equation one obtains the induced 
electron-density linear response, Eq.~(\ref{density2}), which is 
used to get first-order perturbations of the Hartree and 
exchange-correlation potentials. Equation~(\ref{DeltaVBar}) then defines a new 
first-order perturbation of KS potential in the Sternheimer equation. 
This procedure is repeated until the convergence of $\partial n$ is reached.

Within DFPT the responses to perturbations of different wavelengths are decoupled, which 
allows one to calculate vibrational properties at any wavevector without resorting to 
large supercells.\cite{baroni2001phonons} In some cases this can be a strength when compared to other methods such as frozen-phonon (FP)\cite{lam1982ab,togo2023first} or molecular dynamics (MD) methods \cite{mcgaughey2006phonon,kong2011phonon}. 
However, in the case of surfaces, interfaces, or defects, both FP and MD are suitable because the systems naturally require large supercells. 
Another strength of FP is that it is much simpler to implement. On the other hand, {\it{ab initio}} MD explicitly includes temperature dependencies beyond the harmonic approximation, which is particularly important for systems with soft phonons that are dynamically unstable at 0 K.\cite{hellman2011lattice}
Recent developments of machine learning force fields 
have enabled accurate MD simulations with reduced computational cost.\cite{unke2021machine} 
In what follows we focus on calculations within the framework of DFPT.

\section{Charge and thermal transport}
\label{sec:bte}

The study of the dynamics of electrons and phonons in materials, and particularly, thermoelectric effects, belong to the extremely vast field of non-equilibrium statistical physics. 
The macroscopic properties of systems that are not in thermodynamic equilibrium are generally described in terms of their microscopic interactions through either kinetic equations, such as the Kadanoff-Baym \cite{haug2008quantum,stefanucci2013nonequilibrium} or Bloch-Boltzmann formalisms \cite{mahan2010condensed}, 
or by using linear response theory based on the Kubo equations \cite{kubo1957statistical}.
Both approaches consider only near-equilibrium situations with linearized irreversible processes, in which the cornerstone property is the fluctuation-dissipation theorem \cite{kubo1966fluctuation}.
The relation between the Bloch-Boltzmann formalism and Kubo approach was discussed by Thouless \cite{thouless1975relation}, while the derivation of the semiclassical Bloch-Boltzmann formalism from the purely many-body quantum framework of Kadanoff–Baym was given by Ponc{\'e} et.al.~\cite{ponce2020first}. 
The double temporal dependence of the nonequilibrium Green’s functions within Kadanoff-Baym equations, which stems from memory and coherence effects, makes the computational cost of such calculations very demanding. 
Only recently have these equations been implemented using first-principles methods within the completed collision approximation.\cite{sangalli2015ultra}
Such developments may eventually overcome the limitations of semiclassical approaches, however currently first-principles calculations are still mainly based on the semiclassical Bloch-Boltzmann formalism using the Boltzmann Transport Equation (BTE). 

\subsection{Justification for the Bloch-Boltzmann formalism}\label{sec:justification}

The justification for the applicability of the Bloch-Boltzmann formalism to the dynamics of electrons and phonons in materials rests on the concept of quasiparticles within Landau Fermi liquid theory \cite{landau1959theory,pines2018theory}. 
Electrons can be considered to be wavepackets that obey Newtonian laws of motion \cite{pottier2009nonequilibrium}. 
For a wavepacket with well defined momentum, $p=\hbar k$, and hence small uncertainty on the scale of the Fermi momentum, $\Delta k \ll k_F$, the uncertainty principle guarantees that $\Delta x \Delta (\hbar k) \sim \hbar \rightarrow \Delta x \sim \frac{1}{\Delta k} \gg \frac{1}{k_F} \sim a$, where $a$ is the lattice constant. Thus, the typical size of the electron wavepacket is much larger than the lattice constant, $\Delta x \gg a$. 
When this picture is valid, the electron distribution function can be defined over phase space cells larger than $\hbar^3$, the uncertainty principle is not violated, and the BTE can be justified. 
It assumes that classical external fields are varying sufficiently slowly in space and time to justify the use of electron wavepackets.
This approach enables us to relate the transport properties to the Bloch band structure. 

Purely quantum limitations have to be taken into account and 
the validity condition of the BTE can be related to the Peierls inequality \cite{peierls1974some} for non-degenerate 
semiconductors, $\frac{\hbar}{\tau} \ll k_B T$. This means that the time interval in which the 
distribution function evolves should be much smaller than the relaxation time. 
For metals, this condition can be smoothed by considering $\frac{\hbar}{\tau(\epsilon_F)} \ll \epsilon_F$ \cite{peierls1974some},
which can be related to the Mott-Ioffe-Regel criterion  \cite{hussey2004universality}
\begin{equation}
k_F l \gg 1~,
\end{equation}
where $k_F$ is the Fermi wave vector and $l$
is the mean free path of the carriers. 
When this criterion is not fulfilled, such as in 
strongly correlated bad metals \cite{emery1995superconductivity},
the BTE is no longer applicable since the system of interest may not support the existence of long-lived quasiparticles as described by Landau Fermi liquid theory \cite{hartnoll2015theory}. 
Additionally, band transport based on the BTE can fail in describing the case of strong electron-phonon coupling leading to the small polaron limit, in which the charge transport is dominated by thermally activated polaron hopping.\cite{chang2022intermediate} 
Also, another subtle assumption that justifies the BTE is the random phase approximation. 
Within this approximation, the distribution function is only 
given by diagonal terms of the density matrix, while off-diagonal elements are neglected. 
Kohn and Luttinger provided a rigorous mathematical formalism to justify that off-diagonal elements are indeed suppressed by considering an ensemble average over a random distribution of impurities.\cite{kohn1957quantum,luttinger1958quantum}

\subsection{Boltzmann Transport Equation (BTE)}

The Boltzmann Transport Equation describes the propagation of electron 
or phonon distribution functions, $f_{n {\bf{k}}}({\bf{r}},t)$ or $N_{\nu {\bf{q}}}({\bf{r}},t)$, respectively. 
The distribution functions give the probability that an electron (phonon) occupies a state with momentum ${\bf{k}}$ and band $n$ (momentum ${\bf{q}}$ and branch $\nu$) at position ${\bf{r}}$ and time $t$. Within this picture we assume that when the electron (phonon) BTE is solved, the phonon (electron) system remains in equilibrium, which 
relies on the Bloch assumption. Coupled electron-phonon transport that takes into account the drag effect of non-equilibrium electrons (phonons) on the phonons (electrons) involves the solution of a coupled electron–phonon BTE, which can be accomplished 
by a recently released solver called \texttt{elphbolt}.\cite{protik2022elphbolt} In the examples that follow we assume that the drag effect can be ignored.

In the diffusive transport limit and the presence of a temperature gradient ($\nabla T$) and applied electric (${\bf{E}}$) and magnetic fields (${\bf{B}}$),
carrier transport properties can be obtained
by solving the semiclassical BTE for the nonequilibrium
carrier distribution function $f_{n,{\bf{k}}} = f(\epsilon_{n,{\bf{k}}})$
\begin{equation}
\label{boltz1}
\frac{\partial f_{n,{\bf{k}}}}{\partial t} + {\bf{v}}_{n,{\bf{k}}}\cdot\nabla_{{\bf{r}}} f_{n,{\bf{k}}} - \frac{{\bf{F}}}{\hbar}\cdot\nabla_{{\bf{k}}} f_{n,{\bf{k}}} = \left(\frac{\partial f_{n,{\bf{k}}}}{\partial t}\right)_{coll}~. 
\end{equation}
The electronic band velocity of the carrier in the state $\{n,{\bf{k}}\}$ with energy $\epsilon_{n,{\bf{k}}}$
is given by ${\bf{v}}_{n,{\bf{k}}} = \frac{1}{\hbar} \nabla_{{\bf{k}}}\epsilon_{n,{\bf{k}}}$, considering the diagonal matrix elements of the velocity operator and ignoring the Berry curvature.\cite{xiao2010berry} 
The external force is given by ${\bf{F}} = e({\bf{E}} + {\bf{v}}_{n,{\bf{k}}}\times{\bf{B}})$,
where $e$ is the absolute value of the charge of the carriers.
In this equation, the temporal evolution of the electron distribution function results from a balance between 
drift and collision terms. 
The drift term on the left-hand side of Eq.~(\ref{boltz1}) is an external field-driven flow of the space and momentum variables, 
while the collision term on the right-hand side accounts for any relevant scattering mechanisms within the diffusive or hydrodynamic regime.
In Sect.~\ref{sec:ph-ph} we will elaborate on the analogous BTE that can be written for the phonon distribution function, $N_{\nu {\bf{q}}}({\bf{r}},t)$, where the phonons follow the Bose-Einstein distribution at equilibrium and are not subject to the external force, $\bf{F}$.  The electronic and phononic scattering mechanisms will be discussed in Sect.~\ref{sec:scattering-mech}.

The knowledge of $f_{n,{\bf{k}}}$ allows for the evaluation of the charge current density,
\begin{equation}
{\bf{j}} = -\frac{2e}{V}\sum_n\sum_{k}{\bf{v}}_{n,{\bf{k}}}f_{n,{\bf{k}}} = -\frac{2e}{(2\pi)^3}\sum_n\int{{\bf{v}}_{n,{\bf{k}}}f_{n,{\bf{k}}}d{\bf{k}}}~,
\end{equation}
and the heat energy flux density,
\begin{equation}
{\bf{j}}_Q = \frac{2}{V}\sum_n\sum_{{\bf{k}}}\left(\epsilon_{n,{\bf{k}}}-\mu\right){\bf{v}}_{n,{\bf{k}}}f_{n,{\bf{k}}} \\
= \frac{2}{(2\pi)^3}\sum_n\int{\left(\epsilon_{n,{\bf{k}}}-\mu\right){\bf{v}}_{n,{\bf{k}}}f_{n,{\bf{k}}}d{\bf{k}}}~,
\end{equation}
where the factor of 2 appears due to the electron
spin, $V$ is the crystal's volume, and $\mu$ is the chemical potential.

In the next section we discuss the solution of the BTE in the relaxation time approximation for the case of charge carriers; the solution for phonons is analogous. The iterative approach to solving the BTE will be presented in Sect.~\ref{sec:iterative}. 

\subsection{Relaxation Time Approximation (RTA)}
\label{sec:rta}

The collision term on the right-hand side of Eq.~(\ref{boltz1}) drives the system towards a steady state. This scattering term can be expressed
by introducing the per-unit-time probability, $W(n,{\bf{k}}|j,{\bf{{k}^{\prime}}})$, of the
transition of the charge carrier from the state $\{n,{\bf{k}}\}$ to
state $\{j,{\bf{{k}^{\prime}}}\}$, as a result of a particular scattering mechanism.
From the principle of detailed balance, the number of charge carriers
coming into the state $\{j,{\bf{{k}^{\prime}}}\}$ from $\{n,{\bf{k}}\}$ is the same as
the number coming out from $\{j,{\bf{{k}^{\prime}}}\}$ into $\{n,{\bf{k}}\}$. That is, the scattering processes are equally likely going forward or in reverse. This allows us to write
\begin{flalign}
\label{boltz2}
\left(\frac{\partial f_{n,{\bf{k}}}}
{\partial t}\right)_{coll} = \sum_{j,{\bf{{k}^{\prime}}}} [W(j,{\bf{{k}^{\prime}}} | n,{\bf{k}}) f_{j,{\bf{{k}^{\prime}}}} \left(1 - f_{n,{\bf{k}}}\right) 
- W(n,{\bf{k}} | j,{\bf{{k}^{\prime}}}) f_{n,{\bf{k}}} \left(1 - f_{j,{\bf{{k}^{\prime}}}}\right)]~.
\end{flalign}
This condition is a consequence of time-reversal symmetry of the microscopic equations of motion and is sufficient to ensure a positive local entropy production rate for systems out of equilibrium. It is described via the H-theorem\cite{von2010proof} by writing the results as a function of H, the negative of the entropy.\cite{kadanoff2017entropy,allen1996boltzmann} 
The intricate dependency of $f_{n,{\bf{k}}}$ on the linear response distribution function of all other states, $f_{j,{\bf{{k}^{\prime}}}}$, complicates the solution of the BTE and thus various forms of the RTA are usually applied. Ponc{\'e} et.al.~\cite{ponce2020first} reviewed different levels of approximations such as
the momentum relaxation time approximation (MRTA), the self-energy relaxation time approximation (SERTA)\cite{ponce2018towards} and the lowest-order variational approximation (LOVA)\cite{liu2015direct} or the Ziman resistivity formula for metals.\cite{ziman2001electrons}. In general, both the computational cost and the accuracy decreases from the former to the latter. Here we focus on SERTA, which includes forward scattering from $\{n,{\bf{k}}\}$ into $\{j,{\bf{{k}^{\prime}}}\}$ and neglects backward scattering rates into the state $\{n,{\bf{k}}\}$. MRTA partially includes backward scattering rates by considering a geometrical factor that depends on the scattering angle.\cite{ponce2020first}
On the other hand, LOVA includes only an average of the state- and momentum-resolved total decay rates combined with Drude's formula.\cite{Grimvall} 

Assuming that the system is close enough to local equilibrium, 
the non equilibrium distribution function, $f_{n,{\bf{k}}}$, differs only slightly
from that of the equilibrium state, $f_{n,{\bf{k}}}^{(0)}$; namely
$\Delta f(n,{\bf{k}}) = \lvert f_{n,{\bf{k}}}-f_{n,{\bf{k}}}^{(0)}\rvert \ll f_{n,{\bf{k}}}^{(0)}$.
Consequently, $f_{n,{\bf{k}}}$ can be expanded to first order as
\begin{equation}
\label{smallness}
f_{n,{\bf{k}}} = f_{n,{\bf{k}}}^{(0)} -{\tau}_{n,{\bf{k}}}{\bf{v}}_{n,{\bf{k}}}\cdot{\bf{\Phi_0}}(\epsilon)\left(\frac{\partial f^{(0)}}{\partial \epsilon}\right)~,
\end{equation}
where ${\bf{\Phi_0}}(\epsilon) = -e{\mathlarger{\varepsilon}} - \frac{\epsilon - \mu}{T}\nabla T$ is
the generalized disturbing force (dynamic and static) causing the deviation from the
equilibrium distribution, and ${\mathlarger{\varepsilon}} = {\bf{E}} + (1/e)\nabla \mu = -\nabla (\phi_0-(\mu/e))$
is the gradient of the electrochemical potential.
Using this approximation, Eq.~\eqref{boltz2} can be written in the RTA as
\begin{equation}
\label{coll}
\left(\frac{\partial f_{n,{\bf{k}}}}{\partial t}\right)_{coll} = -\frac{\Delta f(n,{\bf{k}})}{\tau_{n,{\bf{k}}}}~, 
\end{equation}
where
\begin{equation}
\label{tau1}
\frac{1}{{\tau_{n,{\bf{k}}}}} = \sum_{{\bf{{k}^{\prime}}}} \sum_j W(n,{\bf{k}}|j,{\bf{{k}^{\prime}}}) \\
\left( \frac{1-f^{(0)}_{j,{\bf{{k}^{\prime}}}}}{1-f^{(0)}_{n,{\bf{k}}}} - \frac{f^{(0)}_{n,{\bf{k}}}}{f^{(0)}_{j,{\bf{{k}^{\prime}}}}}\frac{\Delta f(j,{\bf{{k}^{\prime}}})}{\Delta f(n,{\bf{k}})}\right)~, 
\end{equation}
considering both the absence of quantization effects and that $W(n,{\bf{k}}|j,{\bf{{k}^{\prime}}})$ does not depend on
${\bf{E}}$, ${\bf{B}}$, or $\nabla T$. Substituting Eq.~(\ref{smallness}) into Eq.~(\ref{tau1}) we obtain
\begin{equation}
\label{tau2}
\frac{1}{\tau_{n,{\bf{k}}}} = \sum_{{\bf{{k}^{\prime}}}} \sum_j W(n,{\bf{k}}|j,{\bf{{k}^{\prime}}}) \\
\frac{1-f^{(0)}(\epsilon^{\prime})}{1-f^{(0)}(\epsilon)}\left( 1 - \frac{{\tau}_{j,{\bf{{k}^{\prime}}}}}{\tau_{n,{\bf{k}}}} \frac{{\bf{v}}_{j,{\bf{{k}^{\prime}}}}\cdot\bf{\Phi_0}(\epsilon^{\prime})}{{\bf{v}}_{n,{\bf{k}}}\cdot\bf{\Phi_0}(\epsilon)}\right)~.
\end{equation}
Thus, in the steady-state limit of a homogeneous system with no magnetic field, Eq.~\eqref{boltz1} simplifies to\begin{equation}
\label{boltz3}
{\bf{v}}_{n,{\bf{k}}}\cdot\nabla_{{\bf{r}}} f_{n,{\bf{k}}} - \frac{e {\bf{E}}}{\hbar}\cdot\nabla_{{\bf{k}}} f_{n,{\bf{k}}} = -\frac{\Delta f(n,{\bf{k}})}{\tau_{n,{\bf{k}}}}~, 
\end{equation}
from which the non equilibrium distribution function is obtained provided that $\tau_{n,{\bf{k}}}$ does not depend on
${\bf{E}}$ or $\nabla T$.

A common additional approximation is the assumption that $W(n,{\bf{k}}|j,{\bf{{k}^{\prime}}})$ 
does not depend on ${\bf{k}}$ and ${\bf{{k}^{\prime}}}$ separately, but only on their magnitudes and the angle between them,
$W(n,{\bf{k}}|j,{\bf{{k}^{\prime}}}) = W_{n,j}(\lvert{\bf{k}}\rvert,\lvert{\bf{{k}^{\prime}}}\rvert,{\bf{k}}\cdot{\bf{{k}^{\prime}}})$. 
Then, Eq.~(\ref{tau2})
may be rewritten as~\cite{askerov2009thermodynamics}
\begin{equation}
\label{tau3}
\frac{1}{{\tau_{n,{k}}}} = \sum_{{\bf{{k}^{\prime}}}} \sum_j W(n,{\bf{k}}|j,{\bf{{k}^{\prime}}}) \left(1-\frac{{\bf{k}}\cdot{\bf{{k}^{\prime}}}}{k^2}\right)~, 
\end{equation}
assuming that charge carrier scattering is purely elastic and the dispersion relation is an arbitrary spherically symmetric function of the magnitude of the wavevector (not necessarily parabolic), so that $\epsilon(\lvert{\bf{k}}\rvert)=\epsilon(\lvert{\bf{{k}^{\prime}}}\rvert)$.
Although Eq.~(\ref{tau3}) was derived for isotropic bands, results obtained from it
have been used to study transport properties of  chalcogenides, which are anisotropic~\cite{chaves2021investigating,ahmad2010energy,ravich1971scattering}.
This is possible because the transport properties along the different directions in these materials are mutually independent, except for the magnetoresistance, which critically depends on the anisotropy.

\subsection{Iterative approach for solving the BTE}
\label{sec:iterative}

In some cases it is possible to go beyond the RTA approach to the BTE by using an iterative solution method. 
From perturbation theory, if the interaction potential is weak enough, the scattering can be treated in the Born approximation and Fermi's golden rule can be used to determine the transition probability of the electronic scattering process $\{\ket{n {\bf{k}}}\} \rightarrow \{\ket{m{\bf{k}}+{\bf{q}}}\}$, describing absorption of a phonon with mode $\{\nu\}$ and wave vector ${\bf{q}}$:
\begin{equation}
\label{Fermi}
W_{n {\bf{k}},\nu {\bf{q}}}^{m{\bf{k}}+{\bf{q}}} = \frac{2\pi}{\hbar} |g_{n {\bf{k}},\nu {\bf{q}}}^{m{\bf{k}}+{\bf{q}}}|^2 \delta(\epsilon_{n {\bf{k}}} + \hbar\omega_{\nu {\bf{q}}} - \epsilon_{m{\bf{k}}+{\bf{q}}})~.
\end{equation}
The transition probability exhibits the microreversibility 
property which stems from the time-reversal invariance of the microscopic
equations of motion. Thus the last equation 
is equal to the reverse transition $W^{n {\bf{k}},\nu {\bf{q}}}_{m{\bf{k}}+{\bf{q}}}$ 
from $\{\ket{m{\bf{k}}+{\bf{q}}}\}$ to $\{\ket{n {\bf{k}}}\}$ by emitting a phonon.  
The transition rate at equilibrium, which is the transition per unit time is given by 
\begin{equation}
\label{transition}
\Pi_{n {\bf{k}},\nu {\bf{q}}}^{m{\bf{k}}+{\bf{q}}} = \Pi^{n {\bf{k}},\nu {\bf{q}}}_{m{\bf{k}}+{\bf{q}}} = f_{n {\bf{k}}}^0 (1 - f_{m {\bf{k}} + {\bf{q}}}^0) N_{\nu {\bf{q}}}^0 W_{n {\bf{k}},\nu {\bf{q}}}^{m{\bf{k}}+{\bf{q}}}~.
\end{equation}
By writing the canonical form of the scattering term \cite{ziman2001electrons} of the BTE 
for the case of el-ph coupling 
\begin{equation}
\label{scatt_canon}
\frac{\partial f_{n{\bf{k}}}}{\partial t}\bigg|_{scatt} = - \sum_{\nu {\bf{q}}} \left(\Pi_{n {\bf{k}},\nu {\bf{q}}}^{m{\bf{k}}+{\bf{q}}} + \Pi_{n {\bf{k}}}^{m{\bf{k}}+{\bf{q}}, -\nu {\bf{q}}}\right) \left(\chi_{n {\bf{k}}} - \chi_{m {\bf{k}} + {\bf{q}}} \right)
\end{equation}
where $\chi_{n {\bf{k}}} = \frac{f_{n{\bf{k}}} - f_{n{\bf{k}}}^0}{f_{n{\bf{k}}}^0(1 - f_{n{\bf{k}}}^0)} = \frac{q {\bf{E}}}{k_B T} \cdot {\bf{G}}_{n {\bf{k}}}$ and
\begin{equation}
\label{transition2}
\Pi_{n {\bf{k}}}^{m{\bf{k}}+{\bf{q}}, -\nu {\bf{q}}} = \frac{2\pi}{\hbar} |g_{n {\bf{k}},\nu {\bf{q}}}^{m{\bf{k}}+{\bf{q}}}|^2 f_{n{\bf{k}}}^0 (1 - f_{m{\bf{k}}+{\bf{q}}}^0) (1 + N^0_{-\nu {\bf{q}}})\delta(\epsilon_{n {\bf{k}}} - \hbar\omega_{- \nu {\bf{q}}} - \epsilon_{m{\bf{k}}+{\bf{q}}})~.
\end{equation}

In the steady-state and approximating the drift term by keeping only the linear terms in ${\bf{E}}$, 
the BTE can be linearized as \cite{li2015electrical}
\begin{equation}
\label{linearBTE}
{\bf{v}}_{n {\bf{k}}}f_{n{\bf{k}}}^0(1 - f_{n{\bf{k}}}^0) + \sum_{m, \nu {\bf{q}}} \left(\Pi_{n {\bf{k}},\nu {\bf{q}}}^{m{\bf{k}}+{\bf{q}}} + \Pi_{n {\bf{k}}}^{m{\bf{k}}+{\bf{q}}, -\nu {\bf{q}}} \right){\bf{G}}_{m {\bf{k}} + {\bf{q}}} = {\bf{G}}_{n {\bf{k}}} \sum_{m, \nu {\bf{q}}} \left(\Pi_{n {\bf{k}},\nu {\bf{q}}}^{m{\bf{k}}+{\bf{q}}} + \Pi_{n {\bf{k}}}^{m{\bf{k}}+{\bf{q}}, -\nu {\bf{q}}}\right) ~.
\end{equation}
The relaxation time approximation is reached when the left hand side summation is neglected and then ${\bf{G}}_{n {\bf{k}}}^{\tau} = {\bf{v}}_{n {\bf{k}}}\cdot \tau_{n {\bf{k}}}$, where $\tau_{n {\bf{k}}}$ is given by Eq.~(\ref{imag}). 
Alternatively, Eq.~(\ref{linearBTE}) can be solved by starting with the RTA solution as an initial guess, ${\bf{G}}_{n {\bf{k}}}^{0} = {\bf{G}}_{n {\bf{k}}}^{\tau}$, and iterating using the following relation:
\begin{equation}
\label{iterativeG}
{\bf{G}}_{n {\bf{k}}}^{i+1} = {\bf{G}}_{n {\bf{k}}}^{0} + \frac{\tau_{n {\bf{k}}}}{f_{n{\bf{k}}}^0(1 - f_{n{\bf{k}}}^0)} \left(\Pi_{n {\bf{k}},\nu {\bf{q}}}^{m{\bf{k}}+{\bf{q}}} + \Pi_{n {\bf{k}}}^{m{\bf{k}}+{\bf{q}}, -\nu {\bf{q}}}\right){\bf{G}}^i_{m {\bf{k}} + {\bf{q}}}~.
\end{equation}
This iterative approach has been implemented in \texttt{EPW} \cite{ponce2016epw} for electron transport, \texttt{ShengBTE} \cite{wu2014shengbte} for thermal transport, and for both charge carriers and phonons in \texttt{Perturbo}\cite{zhou2021perturbo} and \texttt{Phoebe}\cite{cepellotti2022phoebe}. 

\subsection{Thermoelectric kinetic coefficients tensors}

The tensorial formalism based on the Onsager-de Groot-Callen model\cite{onsager1931reciprocal,onsager1931reciprocal2,callen1948application,groot1963thermodynamics,callen1985thermodynamics} is appropriate to discuss thermoelectric (TE) effects for anisotropic materials using the thermodynamics of irreversible processes and linear response theory. In this model, the flux of charge carriers and thermal energy is described in terms of a kinetic matrix and generalized forces. (For reviews see Goupil~\cite{goupil2011thermodynamics} and Feldhoff.\cite{feldhoff2015thermoelectric}) 
In summary, the off-diagonal coupling between the electronic current density, ${\bf{j}}$, and
heat energy flux density, ${\bf{j}}_Q$, can be written:
\begin{equation}
\begin{bmatrix} 
    {\bf{j}} \\ 
    {\bf{j}}_Q  
\end{bmatrix}
=
\begin{bmatrix}
    {\bf{L^{11}}} & {\bf{L^{12}}} \\
    {\bf{L^{21}}} & {\bf{L^{22}}}
\end{bmatrix}
\cdot
\begin{bmatrix} 
    {\mathlarger{\varepsilon}} \\
    -\frac{\nabla T}{T}
\end{bmatrix}
\end{equation}
in which, ${\bf{L^{11}}}$, ${\bf{L^{12}}}$, ${\bf{L^{21}}}$,
${\bf{L^{22}}}$ are the moments of the generalized transport
coefficients. 
The Onsager reciprocity relations guarantee ${\bf{L^{12}}} = {\bf{L^{21}}}$ ~\cite{onsager1931reciprocal,onsager1931reciprocal2,callen1985thermodynamics}.
These kinetic coefficients are defined by
\begin{equation}
\label{Lambda}
\Lambda^{(\alpha)}(\mu;T)= e{^2}\int\Xi(\epsilon,\mu,T)(\epsilon - \mu)^{\alpha}\left(-\frac{\partial f^{(0)}(\mu;\epsilon,T)}{\partial \epsilon}\right)d\epsilon~, 
\end{equation}
with ${\bf{L^{11}}} = \Lambda^{(0)}$,
${\bf{L^{21}}} = {\bf{L^{12}}} = -(1/e)\Lambda^{(1)}$, and
${\bf{L^{22}}} = (1/e^2)\Lambda^{(2)}$, in which $\Xi(\epsilon,\mu,T)$
is the transport distribution kernel (TDK) given by
\begin{equation}
\Xi(\epsilon,\mu,T) = \int \sum_n{{\bf{v}}_{n,{\bf{k}}}\otimes{\bf{v}}_{n,{\bf{k}}}{\tau}_{n,{k}}}(\mu,T)\delta(\epsilon - \epsilon_{n,{\bf{k}}})\frac{d{\bf{k}}}{8\pi^3}~. 
\end{equation}

In the dynamic steady state, thermoelectric properties can be obtained by considering specific experimental conditions. 
For an isothermal situation ($\nabla T = 0$), the charge  current, ${\bf{j}}$, obeys Ohm's law and the kinetic coefficient tensor
can be identified with the electrical conductivity tensor, $\sigma = \Lambda^{(0)}$.
As pointed out by Feldhoff~\cite{feldhoff2015thermoelectric}, 
even with the isothermal condition, ${\bf{j}}$ is accompanied by an entropy current whose magnitude and direction depend on the Seebeck coefficient, $S$, which can be viewed as a quantity that measures entropy flow per unit charge. 
$S$ is obtained by requiring the electric current to vanish, so the electrochemical potential gradient and the thermal gradient are balanced. Then $S$ is given by the ratio between them, $S = (eT)^{-1}\Lambda^{(1)}/\Lambda^{(0)}$. 

As one of the most sensitive probes of the carriers in a material, the Seebeck measurement is related to the heat per carrier over temperature or the entropy per carrier, as pointed out earlier. It suggests that $S$ can provide information about $i)$ the sign of the charge of the carriers and $ii)$ the characteristic energy associated with the carriers. However, the former result does not hold in a few cases, such as noble metals, where $S$ and the Hall coefficient have diverging signs.\cite{chaikin1990introduction} This has been attributed to the complex energy dependence of the mean free path due to the electron-phonon scattering.\cite{robinson1967thermoelectric}. The second result generally holds and can be used, for example, to distinguish metals and semiconductors by comparing the magnitude and temperature behavior of $S$. While metals exhibit values of $S$ that decrease with temperature and have magnitudes much smaller than $k_B/e \approx 87 \mu V/K$, semiconductors present much larger magnitudes that increase with temperature.\cite{chaikin1990introduction} 

The zero electric current condition yields Fourier’s law,
in which the charge carrier contribution to the thermal conductivity tensor is given by 
\begin{equation}
\kappa_{el} = (e^2T)^{-1} \left({\Lambda^{(1)}\cdot{\Lambda^{(0)}}^{-1}}\cdot{\Lambda^{(1)}} - \Lambda^{(2)}\right)~. 
\end{equation}
The second term within brackets corresponds to the thermal conductivity due to the carrier transport under isoelectrochemical conditions, while the first term is the power factor ($PF=\sigma S^2$) related to the thermal conductivity that couples to the charge current. 
This is responsible for thermoelectric conversion, that is, the transfer of energy from an entropy current to an electric current or vice versa.\cite{feldhoff2014high,feldhoff2015thermoelectric}

\section{Scattering mechanisms within semiclassical BTE}
\label{sec:scattering-mech}

In order to solve the BTE for either charge carriers or phonons we need to provide the collision term on the right-hand side of Eq.~(\ref{boltz1}). This term incorporates the various microscopic scattering mechanisms that are detailed below.

\subsection{Electron-phonon interaction}
\label{sec:el-ph}

\begin{figure}[t]
\includegraphics[width=0.85\columnwidth]{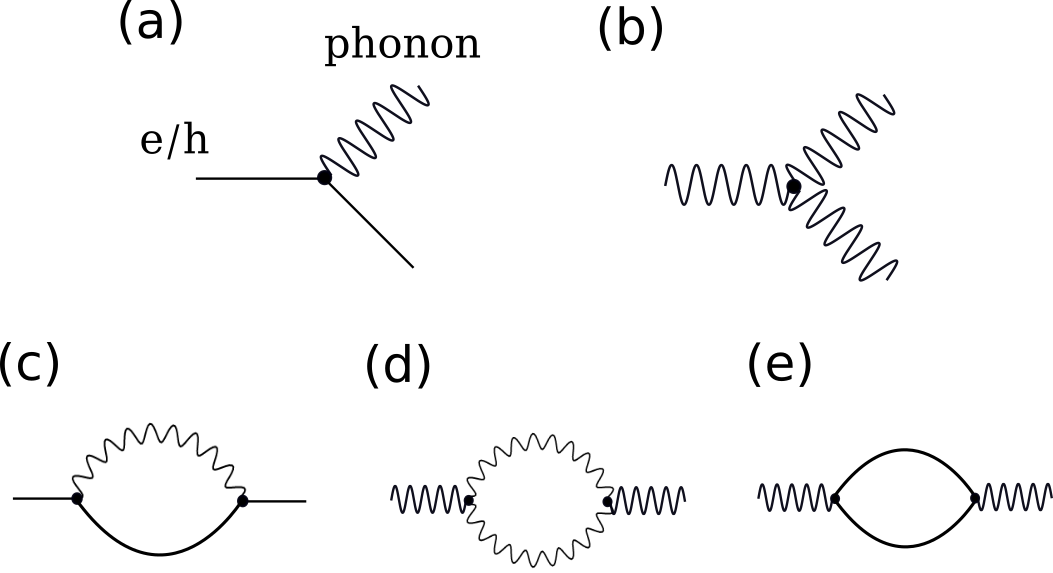}
\caption{\label{fig:diagrams} Feynman diagrams showing (a) electron-phonon and (b) three-phonon interactions, as well as the self-energy diagrams with contributions from (c) electron-phonon, (d) phonon-phonon, and (e) phonon-electron scattering.} 
\end{figure}

The electron–phonon interaction is a key factor in determining functional properties of materials such as thermoelectric transport properties.
The electron-phonon coupling hamiltonian can be derived within DFT by expanding 
the KS Hamiltonian to first-order in $\Delta \xi_{\kappa p}$, the displacements of the nuclei from their equilibrium positions, $\xi^0_{\kappa p}$:
\begin{equation}
\label{expandKS}
V^{KS}(\{\xi_{\kappa p} \}) = V^{KS}(\{\xi^0_{\kappa p} \}) + \sum_{\kappa \alpha p} \frac{\partial V^{KS}}{\partial \xi_{\kappa \alpha p}} \Delta \xi_{\kappa \alpha p}~.
\end{equation}
Extending the expansion to second-order results in Debye-Waller terms \cite{antonvcik1955theory} that will not be discussed in the present review, 
since these terms are purely real and do not affect el-ph relaxation times.
\cite{lautenschlager1986phonon}
The operator for the phonon perturbation potential is given in terms of quantized 
normal mode coordinates using phonon annihilation and creation operators ($b_{\nu\mathbf{q}},b^\dag_{\nu\mathbf{q}}$):
\begin{equation}
\label{expandKSq}
V^{KS} = V^{KS}(\{\xi^0_{\kappa p} \}) + \frac{1}{N_p^{1/2}}\sum_{\nu {\bf{q}}} \Delta_{\nu {\bf{q}}} V^{KS} (\hat{b}_{\nu {\bf{q}}} + \hat{b}^{\dagger}_{-\nu {\bf{q}}})~, 
\end{equation}
where 
\begin{equation}
\label{Delta1}
\Delta_{\nu {\bf{q}}} V^{KS} = e^{i{\bf{q}}\cdot{\bf{r}}} \Delta_{\nu {\bf{q}}} v^{KS}~,
\end{equation}
and 
\begin{equation}
\label{Delta2}
\Delta_{\nu {\bf{q}}} v^{KS} = c_{\nu {\bf{q}}} \sum_{\kappa \alpha p} \left(\frac{M_0}{M_{\kappa}}\right)^{\frac{1}{2}} e_{\kappa \alpha, \nu}({\bf{q}}) e^{-i{\bf{q}}\cdot({\bf{r}} - {\bf{R}}_p)} \frac{\partial V^{KS}}{\partial \xi_{\kappa \alpha }}\bigg|_{{\bf{r}} - {\bf{R}}_p} ~,
\end{equation}
with $c_{\nu {\bf{q}}}$ being the zero-point displacement amplitude.
Using the electron annihilation and creation operators, ($a_{n\mathrm{k}}, a^\dag_{n\mathrm{k}}$), the first-principles el-ph perturbation Hamiltonian is
\begin{eqnarray}
\label{eq:Helph}
\hat{H}^{el-ph} &=& \sum_{n{\bf{k}},n^{\prime}{\bf{k^{\prime}}}} \braket{\phi_{n^{\prime}{\bf{k^{\prime}}}}|V^{KS}(\{\xi_{\kappa p} \})- V^{KS}(\{\xi^0_{\kappa p} \})|\phi_{n{\bf{k}}}} \hat{b}_{\nu {\bf{q}}} \hat{a}^{\dagger}_{n {\bf{k}}} \hat{a}_{n^{\prime} {\bf{k^{\prime}}}} \nonumber \\
&=& \frac{1}{N_p^{1/2}} \sum_{{\bf{k}},{\bf{q}}m n \nu} g_{mn\nu}({\bf{k}},{\bf{q}}) \hat{a}^{\dagger}_{n {\bf{k}}} \hat{a}_{n^{\prime} {\bf{k^{\prime}}}} (\hat{b}_{\nu {\bf{q}}} + \hat{b}^{\dagger}_{-\nu {\bf{q}}})~, 
\end{eqnarray}
where the electron-phonon matrix element has been defined as
\begin{equation}
\label{gkq}
g_{mn\nu}({\bf{k}},{\bf{q}}) = \braket{\psi_{m\mathbf{k}+ \mathbf{q}}|\Delta_{\nu {\bf{q}}} v^{KS}|\psi_{n{\bf{k}}}}~,
\end{equation}
which can be computed by DFPT (see Sect.~\ref{sec:dfpt}), using Eq.~(\ref{density2}) \cite{giustino2017electron}.
From the electron-phonon Hamiltonian, the el-ph relaxation times (RT) can be derived in different ways, such as perturbatively via Feynman-Dyson diagram techniques \cite{keating1968dielectric,marini2015many} or  nonperturbatively via Hedin-Baym equations within quantum field theory  \cite{baym1961field,hedin1970effects,giustino2017electron}.
Also, within the Born approximation, the RT can be derived by considering 
all scatterings in which an electron emits or absorbs one phonon via Fermi's golden rule, 
along with the rate equation for the time-dependent electron distribution functions \cite{Grimvall}.
The latter approach was derived in Section~\ref{sec:iterative}. In all cases the RT is given by the imaginary part of the electron self energy, $\Sigma$: 
\begin{equation}
\label{imag}
 \begin{split}
\operatorname{Im} \left[ \Sigma_{n,{\bf{k}}}(\epsilon,T) \right] =\pi \sum_{m,\nu} \int_{BZ} \frac{d{\bf{q}}}{\Omega_{BZ}} |g_{mn\nu}({\bf{k},{\bf{q}}})|^2 \\
       \times \Bigg[\left[N_{\nu{\bf{q}}}(T) + f_{m{\bf{k}}+{\bf{q}}}\right]\delta(\epsilon - (\epsilon_{m{\bf{k}}+{\bf{q}}} - \epsilon_F) + \hbar\omega_{{\bf{q}}\nu}) \\
        + [N_{\nu{\bf{q}}}(T) + 1 - f_{m{\bf{k}}+{\bf{q}}}]\delta(\epsilon - (\epsilon_{m{\bf{k}}+{\bf{q}}} - \epsilon_F) - \hbar\omega_{{\bf{q}}\nu})\Bigg]~,
\end{split}
\end{equation}
where $g_{mn\nu}({\bf{k},{\bf{q}}})$ are the 
el-ph matrix elements, $\Omega_{BZ}$ is the volume of the Brillouin zone (BZ) and the Dirac $\delta$-functions enforce energy conservation for emission or absorption of a phonon with wavevector ${\bf{q}}$ and mode $\nu$ and energy $\hbar \omega_{\nu {\bf{q}}}$. 
Equation~(\ref{imag}) contains the dynamical structure of the electron-phonon interaction on the scale 
of the phonon energy. 
The temperature dependence comes from the electron and phonon distribution functions. 
Eq.~(\ref{imag}) corresponds to the imaginary part of the 
lowest-order Feynman diagram for the electron self-energy shown in Fig.~\ref{fig:diagrams}(c).
Migdal theory demonstrated that non-adiabatic terms are less important in higher order diagrams, justifying the truncation at the lowest order diagram \cite{migdal1958interaction,allen1983theory}.

Since the RTs are resolved for different bands and ${\bf{k}}$-points, after integrating over all phonon modes and ${\bf{q}}$-points, first-principles calculations can provide rich microscopic information. 
In order to converge the el-ph RTs when calculating transport properties, 
very dense ${\bf{k}}$ and ${\bf{q}}$ meshes are needed \cite{li2015electrical,mustafa2016ab}, so interpolation schemes have to be used instead brute-force DFPT calculations. In Sect.~\ref{sec:interpolation} we discuss interpolation schemes developed for this aim. 

\subsection{Scattering of electrons}

\subsubsection{Fr{\"o}hlich dipole and quadrupolar el-ph coupling} 
        
While screening lengths are very short in metals, long-range (LR) el-ph interactions may appear in semiconductors and insulators due to the
incomplete screening of the potential generated by the atomic displacements.
In particular, polar materials, with two or more atoms in the unit cell,
exhibit nonzero Born effective charge tensors \cite{gonze1997dynamical,verdi2015frohlich}.
In these systems, the most relevant contribution to the
perturbation potential ($\partial V^{KS}/\partial \xi_{\kappa\alpha}$) is a dipole
decaying as $|{\bf{R}}_p^{-2}|$, and thus long-range el-ph interactions occur in the long-wavelength
limit $({\bf{q}} \to 0)$, and are
responsible for the longitudinal optical-transverse optical (LO-TO) splitting
of the optical frequencies \cite{born1954k} and the divergence of the el-ph
matrix elements, known as Fr{\"o}hlich coupling \cite{frohlich1937electrical}.
This polar scattering was first discussed
by Fr{\"o}hlich~\cite{frohlich1937electrical}, Callen~\cite{callen1949electric}, and Howarth and Sondheimer~\cite{howarth1953theory}.

In polar materials lacking inversion symmetry the piezoelectric (PE) el-ph interaction
plays an important role, in addition to the dipolar and deformation-potential contributions.
Such interactions come from the strain induced by acoustic phonons and can be
expressed as a function of the macroscopic piezoelectric constants of the material \cite{mahan2010condensed}.
In a more general approach, all these el-ph interactions, including dipolar, deformation-potential and
piezoelectric contributions, can be expressed as a multipole Vogl expansion of the
el-ph potential \cite{vogl1976microscopic,lawaetz1969long}.
The divergent Fr{\"o}hlich el-ph interaction comes from the dipole term, while
the piezoelectric interaction comes from both the dipole and the quadrupole terms \cite{vogl1976microscopic}.
Even though the Fr{\"o}hlich coupling is dominant for LO modes \cite{verdi2015frohlich},
quadrupolar interactions can dominate for the TO and acoustic modes \cite{vogl1976microscopic}.

Because Wannier-Fourier (WF) interpolation (described in Sect.~\ref{sec:interpolation}) is based on the spatial localization of the el-ph coupling,
first-principles treatment of the LR interaction is not amenable
to WF interpolation, since it requires a very large number
of e-p matrix elements to attain convergence \cite{rohlfing2000electron}.
Verdi \cite{verdi2015frohlich} and Sjakste et al. \cite{sjakste2015wannier} developed a first-principles approach to
adapt the WF interpolation to the case of dipole Fr{\"o}hlich coupling in polar materials, and the importance of the next-to-leading order term in the Vogl expansion beyond the dipolar contribution was addressed by Brunin \cite{brunin2020electron,brunin2020phonon}, Jhalani \cite{jhalani2020piezoelectric} and Park \cite{park2020long}.
In summary, those works separate the LR dipole and quadrupole
contributions to the el-ph matrix elements from the short-range (SR) part:
\begin{eqnarray}
\label{splitg}
g_{mn\nu}({\bf{k}},{\bf{q}}) &=& g_{mn\nu}^L({\bf{k}},{\bf{q}}) + g_{mn\nu}^S({\bf{k}},{\bf{q}}) \\
                             &=& g_{mn\nu}^D({\bf{k}},{\bf{q}}) + g_{mn\nu}^Q({\bf{k}},{\bf{q}}) + g_{mn\nu}^S({\bf{k}},{\bf{q}})~,
\end{eqnarray}
where $g_{mn\nu}^D({\bf{k}},{\bf{q}})$ is the first-principles Fr{\"o}hlich el-ph matrix element,
written in terms of Born effective charges \cite{verdi2015frohlich}.
\begin{multline}
\label{long}
g_{mn,\nu}^D({\bf{k}},{\bf{q}}) = i \frac{e^2}{\Omega_{uc}\epsilon_0} \sum_{\kappa}
 \left(\frac{\hslash}{2 N_p{M_{\kappa} \omega_{{\bf {q}}\nu}}}\right)^{\!\!\frac{1}{2}} \\
 \times \sum_{{\bf G}\ne -{\bf q}} 
 \frac{ ({\bf {q}}+{\bf {G}})\cdot{\bf {Z}}^*_\kappa \cdot {\bf {e}}_{\kappa\nu}({\bf {q}}) } 
 {({\bf {q}}+{\bf {G}})\cdot\bm\zeta_\infty\!\cdot({\bf {q}}+{\bf {G}})} \\
 \times \braket{\Psi_{m{\bf{k}}+{\bf{q}}}|e^{i({\bf{k}}+{\bf{q}})\cdot{\bf{r}}}|\Psi_{n{\bf{k}}}}~, 
\end{multline}
in which
$N_p$ is the number of unit cells in the Born-von K{\'a}rm{\'a}n supercell, $\Omega_{uc}$ is the volume of the unit cell, 
${\bf{G}}$ is a reciprocal lattice vector, ${\bf{Z}}^*=Z^*_{\alpha,\beta}$
is the Born effective charge tensor, ${\bf {e}}_{\kappa\nu}({\bf {q}})$
is a phonon eigenmode normalized within the unit cell, $\bm\zeta_\infty = \zeta^{\infty}_{\alpha,\beta}$
corresponds to the high-frequency dielectric constant tensor, $\epsilon_0$ is the vacuum permittivity, and $\hslash$
is the reduced Planck constant.
$\braket{\Psi_{m{\bf{k}}+{\bf{q}}}|e^{i({\bf{k}}+{\bf{q}})\cdot{\bf{r}}}|\Psi_{n{\bf{k}}}} = \left[ U^{{\bf {k}}+{\bf {q}}}\:U^{{\bf {k \dagger}}} \right]_{mn}$ are phase factors given in terms of rotation matrices, $U^{{\bf {k}}+{\bf {q}}}$,
that appear in the definition of the maximally localized Wannier functions (MLWFs), Eq.~(\ref{eqn:mlwf})\cite{marzari1997maximally}.

The quadrupole el-ph matrix elements imply summations over the  Cartesian indices $\alpha$, $\beta$ and $\gamma$ and can be expressed as \cite{jhalani2020piezoelectric,park2020long}
\begin{multline}
\label{long2}
g_{mn,\nu}^Q({\bf{k}},{\bf{q}}) = \frac{e^2}{\Omega_{uc}\epsilon_0} \sum_{\kappa}
 \left(\frac{\hslash}{2 {M_{\kappa} \omega_{{\bf {q}}\nu}}}\right)^{\!\!\frac{1}{2}} \\
 \times \sum_{{\bf G}\ne -{\bf q}} 
 \frac{ \frac{1}{2}(q_{\alpha}+G_{\alpha})(Q_{\kappa,\beta}^{\alpha\gamma}{\bf{e}}^{(\beta)}_{\kappa \nu} ({\bf{q}}))(q_{\gamma}+G_{\gamma})}{(q_{\alpha}+G_{\alpha})\zeta_{\alpha\gamma}(q_{\gamma}+G_{\gamma}) }   \\
 \times \braket{\Psi_{m{\bf{k}}+{\bf{q}}}|e^{i({\bf{q}}+{\bf{G}})\cdot({{\bf{r}} - \xi_{\kappa}})}|\Psi_{n{\bf{k}}}}~, 
\end{multline}
where
$Q_{\kappa,\beta}^{\alpha\gamma}$ are the dynamical quadrupoles that
correspond to the second order term of a multipole
expansion of the charge density induced by an atomic
displacement in the long-wavelength limit, that can also be calculated using the DFPT approach \cite{martin1972piezoelectricity,stengel2013flexoelectricity,dreyer2018current,royo2019first,ponce2021first}. 
As shown for GaN and PbTiO$_3$ \cite{jhalani2020piezoelectric,park2020long},
the inclusion of dipole and quadrupole interactions corrects the long-range el-ph coupling in these polar materials. 
Also, the application of this approach to Si has demonstrated that quadrupole interactions develop an important role even for nonpolar materials \cite{park2020long}.      
            
\subsubsection{Screening for long-range interactions}
            
The above expressions represent a first-principles generalization of the Fr{\"o}hlich and quadrupolar el-ph coupling.
However, this theory is limited to undoped systems, neglecting
screening effects due to the presence of free carriers.
The effect of this additional screening is related to the variation of the
dielectric properties in the long-wavelenth limit and it was discussed recently by Ren et al. \cite{ren2020establishing}
on the basis of the semi-empirical Thomas-Fermi formalism applied to doped half-Heusler semiconductors.
Also, on the basis of a linear-response and dielectric matrix formulation \cite{pick1970microscopic,vogl1976microscopic,stengel2013flexoelectricity},
Macheda et al. \cite{macheda2022electron} developed a first-principles
framework to take into account the screening effects due to the presence of free carriers in doped semiconductors at finite temperature.
In general, free carriers screen out the electric field, resulting in both a weakening of the LR e-p coupling as more carriers are added to the system, and a shift of the frequency of the LO mode~\cite{ehrenreich1959screening} leading into a reduction of the LO-TO splitting.
Neglecting screening effects clearly yields an overestimation of LR el-ph relaxation times and mobility.

A quantification of those effects was addressed within a quasi-static approximation by Ehrenreich~\cite{ehrenreich1959screening}, in which the LR el-ph matrix element is weakened by a factor of $1-(r_{\infty}{\bf{q}})^{-2}$, where $r_{\infty}$ is the screening radius,
\begin{equation}
\label{r0inf}
r{_{\infty}}^{-2}(n,{\bf{k}}) = \frac{4\pi{e^2}}{\zeta{_{\infty}}}\int{\left(-\frac{\partial f_{\mu}(T,\epsilon)}{\partial \epsilon_{n,{\bf{k}}}}\right) g(\epsilon) d\epsilon}~,
\end{equation}
and $g(\epsilon)$ is the density of states (DOS),
\begin{equation}
\label{dos2}
g(\epsilon) = \int \sum_n \delta(\epsilon - \epsilon_{n,{\bf{k}}}) \frac{d{\bf{k}}}{8\pi^3} = \frac{1}{\Omega_{BZ} \mathcal{N}_{{\bf{k}}}}\sum_{n,{\bf{k}}} \frac{\delta(\epsilon - \epsilon_{n,{\bf{k}}})}{d\epsilon}~,
\end{equation}
where $\mathcal{N}_k$ is the number of ${\bf{k}}$-points. 
The eigenfrequency shift of the
LO phonons reads
\begin{equation}
\label{freq_shift}
{(\omega^{LO}})^2 = {(\omega^{TO}})^2\left(\frac{\zeta_0/\zeta_{\infty}+ (r_{\infty}{\bf{q}})^{-2}}{1+(r_{\infty}{\bf{q}})^{-2}}\right)~,
\end{equation}
where $\omega^{TO}$ is the TO mode eigenfrequency.
The eigenfrequency of the LO vibration
is strongly reduced, further altering the e-p matrix elements~\cite{ravich1971scattering}.
The resulting change in the LR RT is given by the following band-dependent factor:
\begin{equation}
\label{screen_pol}
 \begin{split}
& F_\mathrm{pol}(n,{\bf{k}}) = \left[1 -\frac{1}{2(r_{\infty}(n,{\bf{k}})\cdot{\bf{k}})^2}\right.
\\
& \left.\times \ln[1+4(r_{\infty}(n,{\bf{k}})\cdot{\bf{k}})^2] + \frac{1}{1+4(r_{\infty}(n,{\bf{k}})\cdot{\bf{k}})^2}\right]^{-1}~.
\end{split} 
\end{equation}

By combining Eqs.~(\ref{imag}), (\ref{long}), and (\ref{screen_pol}), we arrive at expressions for the RT corresponding to both
non-polar ($\tau_\mathrm{npol}$) and screened polar ($\tau_\mathrm{pol}$) phonon scattering.
The non-polar e-p RT is given by
\begin{equation}
\label{npol}
\frac{1}{\tau_\mathrm{npol}(n,{\bf{k}})} = 2\, \mathrm{Im\,} \Sigma_{n,{\bf{k}}}[\epsilon = \epsilon_{n,{\bf{k}}} - \epsilon_F,T,g^{{S}}_{mn,\nu}({\bf{k},{\bf{q}}})]~,
\end{equation}
and the screened polar e-p RT reads
\begin{equation}
\label{pol}
\frac{1}{\tau_\mathrm{pol}(n,{\bf{k}})} = 2\, \mathrm{Im\,} \Sigma_{n,{\bf{k}}}[\epsilon = \epsilon_{n,{\bf{k}}} - \epsilon_F,T,g^{{L}}_{mn,\nu}({\bf{k},{\bf{q}}})]\times F_\mathrm{pol}(n,{\bf{k}})~.
\end{equation}
It is important to note that dynamical features of the screening were neglected here, since their effect is
regarded  to be quite small~\cite{ravich1971scattering}.
Also, the energy dependence of the RT is also changed because of the energy dependence of the screening that enters through $r_{\infty}$.
This quasi-static approach has been applied recently to address the screened Fr{\"o}hlich coupling in
thermoelectric layered materials \cite{chaves2021microscopic,chaves2022out} and will be reviewed later in this article.
            
\subsubsection{Defect scattering}
            
Carrier scattering by defects is the dominant scattering mechanism that limits charge and spin transport  
in non-degenerate semiconductors at low temperature \cite{radisavljevic2013mobility,li2016charge}.
In this regime subtle quantum transport effects can be induced by defects.\cite{bergmann1984weak,lee1985universal,datta1997electronic} 
Even at higher temperatures, carrier dynamics and thermoelectric properties in highly doped materials can be limited by electron-defect (e-d) scattering. A prominent example is SnSe, in which there is temperature-induced Sn vacancy formation above 600~K \cite{dewandre2016two,chaves2021microscopic}.
As defects are systematically employed to engineer advanced functional materials and devices \cite{gunlycke2011graphene,koenraad2011single}, in particular, thermoelectric materials \cite{zheng2021defect},
the microscopic understanding of e-d scattering from first-principles calculations provides a basis to explore charge and spin dynamics in materials in the presence of neutral and ionized defects.

Calculations of e-d scattering have mostly relied on semiempirical models.
In particular, ionized defect scattering has been treated theoretically by Brooks and Herring (B-H)~\cite{brooks1955theory,chattopadhyay1981electron}.
They used a screened Coulomb potential within the
Born approximation for the evaluation of transition probabilities due to the scattering of carriers by dilute concentrations randomly distributed, ionized scattering centers. 
The dilute regime allows one to neglect the perturbations to the electron energy levels as well as complex effects such as the contributions from coherent scattering off pairs of defect centers, which requires a quantum transport theory~\cite{moore1967quantum}. 
The per-unit-time transition probability for the scattering of charge carriers by ionized defects can be written in the plane-wave approximation as
\begin{equation}
\label{U}
W({\bf{k}}|{\bf{{k}^{\prime}}}) = \frac{2\pi}{\hbar}\frac{N_i}{V} \\
\left\vert{\int U({\bf{r}}) \exp\left[i({\bf{k}}-{\bf{{k}^{\prime}}})\cdot{\bf{r}}\right]d{\bf{r}}}\right\vert^2 \delta(\epsilon_{{\bf{{k}^{\prime}}}}-\epsilon_{{\bf{k}}})~,
\end{equation}
where $U({\bf{r}})$ is the scattering potential and $N_i$ is the number of ionized defects. 
The straightforward application of the long-range Coulomb field,
$U({\bf{r}}) = e\phi(\mathbf{r}) = \pm e^2/\zeta{_0} r$, with electrostatic potential $\phi$ due to the presence of positive (donor) or negative (acceptor) defect ions, leads 
into a logarithmic divergence. Hence a screened Coulomb potential such as
\begin{equation}
U({\bf{r}}) = \pm \frac{e^2}{\zeta{_0} r} \left(e^{-r/r_0}\right)~,
\end{equation}
must be considered, where $r_0$ is the screening radius of the defect ion defined by 
\begin{equation}
\label{r0}
r{_0}^{-2}(k) = \frac{4\pi{e^2}}{\zeta{_0}}\int{-\frac{\partial f^{(0)}(\epsilon)}{\partial \epsilon_k}}g(\epsilon)d\epsilon~. 
\end{equation}
From Eq.~(\ref{U}) and Fermi`s golden rule, the RT for the scattering of
charge carriers by ionized defects can be expressed as \cite{chaves2021investigating}
\begin{equation}
\label{tau_imp}
\tau_\mathrm{imp}(k) = \frac{\hbar\zeta{_0}{^2}}{2{\pi}{e^4}{N_i}F_\mathrm{imp}(k)}k^2 \left\vert\frac{\partial \epsilon_k}{\partial k}\right\vert 
\end{equation}
where
\begin{equation}
F_\mathrm{imp}(k) = \ln(1+\eta) - \frac{\eta}{1+\eta}~, 
\end{equation}
is the screening function, with $\eta = (2kr_0)^2$.

More intricate \emph{ab initio} calculations based on themultiple scattering Korringa-Kohn-Rostoker (KKR) Green’s function method have also been commonly employed \cite{papanikolaou1997lattice,settels1999ab,hohler2004cd,ebert2011calculating}, even though this method is much more computationally expensive. 
In general, those calculations start with DFT, in which 
the Green's functions of a crystal host with a single defect are exactly embedded in the unperturbed crystal host using a Dyson equation.
On the other hand, first-principles calculations of e–d scattering on the basis of 
plane-wave DFT using pseudopotentials or projector augmented waves \cite{restrepo2009first,lordi2010charge} have faced computational challenges due to the high costs of the supercell approach to determine e-d interaction matrix elements within a perturbative approach. 
One recently developed method, based only on the primitive cell and Wannier-Fourier interpolation, 
significantly reduces the computational cost \cite{lu2019efficient,lu2020ab}. 
However, this method was developed only for neutral defects. 
Since the concentration of ionized defects is usually considerably larger than that of neutral imperfections,\cite{chaves2021investigating} an extension of this method to include ionized defects would be necessary.

\subsection{Scattering of phonons}
\label{sec:ph-ph}

Phonon properties can be determined perturbatively by expanding the ionic potential energy in a Taylor series in atomic displacements, as discussed in Sect.~\ref{sec:dfpt}.
The first-derivative term vanishes for a crystal in equilibrium. The second derivative term gives rise to the 2nd-order force constants which describe the phonon band structure through the dynamical matrix in the ``harmonic approximation", Eq.~(\ref{secular}). The third derivative term,
\begin{equation}
\label{eq:3rd-order}
    \frac{\partial^3 E({{\bf{R}}})}{\partial {\bf{R}}_I\partial {\bf{R}}_J \partial {\bf{R}}_K} \ ,
\end{equation}
represents the coupling of three phonons, Fig.~\ref{fig:diagrams}(b), and is the first to allow for scattering between phonons. Diagrammatically, the phonon-phonon scattering rate is related to the imaginary part of the phonon self-energy contributed by a virtual phonon-phonon pair, as shown in Fig.~\ref{fig:diagrams}(d). The three-phonon coupling gives contributions to the phonon scattering matrix arising from phonon absorption (two phonons merge into one) and decay (one phonon splits into two). 

The phonon BTE represents a balance between diffusion due to a temperature gradient and scattering due to various processes. Adapting Eq.~(\ref{boltz1}) to phonons, and following the approach in Ref.~\citenum{fugallo2013ab}, the phonon BTE for the perturbed phonon distribution function $N_{\nu\mathrm{q}}$ can be written
\begin{equation}
    -\mathbf{v}_{\nu\mathbf{q}} \frac{\partial T}{\partial x} \left( \frac{\partial N_{\nu\mathbf{q}}}{\partial T} \right) + \left. \frac{\partial N_{\nu\mathbf{q}}}{\partial t} \right|_\mathrm{scatt} = 0,
\end{equation}
where $\mathbf{v}_{\nu\mathbf{q}}$ is the phonon group velocity. Expanding around the equilibrium Bose-Einstein phonon distribution, $\bar{N}_{\nu\mathrm{q}}$,
\begin{equation}
    N_{\nu\mathrm{q}} = \bar{N}_{\nu\mathrm{q}} + \bar{N}_{\nu\mathrm{q}} (\bar{N}_{\nu\mathrm{q}}+1)\frac{\partial T}{\partial x} \delta N_{\nu\mathrm{q}},
\end{equation}
the BTE can be linearized and written
\begin{eqnarray}
        -\mathbf{v}_{\nu\mathbf{q}} \left( \frac{\partial N_{\nu\mathbf{q}}}{\partial T} \right) &=&
     \sum_{\nu'\mathbf{q'},\nu''\mathbf{q''}} \left[ P^{\nu''\mathbf{q''}}_{\nu\mathbf{q},\nu'\mathbf{q'}} \left( \delta N_{\nu\mathbf{q}} + \delta N_{\nu'\mathbf{q'}} - \delta N_{\nu''\mathbf{q''}} \right) \right. \label{eq:ph1} \\
    &&  \left. +\frac{1}{2} P_{\nu\mathbf{q}}^{\nu'\mathbf{q'},\nu''\mathbf{q''}} \left( \delta N_{\nu\mathbf{q}} - \delta N_{\nu'\mathbf{q'}} - \delta N_{\nu''\mathbf{q''}} \right) \right] \label{eq:ph2} \\
    && +\sum_{\nu'\mathbf{q'}} P^\mathrm{isotope}_{\nu\mathbf{q},\nu'\mathbf{q'}} 
    \left( \delta N_{\nu\mathbf{q}} -\delta N_{\nu'\mathbf{q'}}\right) \label{eq:iso} \\
    && + P^\mathrm{boundary}_{\nu\mathbf{q}} \delta N_{\nu\mathbf{q}} + P^\mathrm{ph-el}_{\nu\mathbf{q}} \delta N_{\nu\mathbf{q}}. \label{eq:ph-el}
    \end{eqnarray}

\subsubsection{Phonon-phonon scattering}
\label{subsec:ph-ph}

The various $P$ matrices encode scattering between different phonon states. In particular, the phonon-phonon scattering terms in Eqs.~(\ref{eq:ph1}) and (\ref{eq:ph2}),
\begin{eqnarray}
    P^{\nu''\mathbf{q''}}_{\nu\mathbf{q},\nu'\mathbf{q'}} &=& \frac{2\pi}{\mathcal{N}_q\hbar^2} \sum_\mathbf{G} \left|V^{(3)}(\nu\mathbf{q},\nu'\mathbf{q'},\nu''\mathbf{-q''}) \right|^2 \nonumber \\
&&    \times \bar{N}_{\nu\mathbf{q}} \bar{N}_{\nu'\mathbf{q'}} (\bar{N}_{\nu''\mathbf{q''}}+1) \delta_{\mathbf{q+q'-q'',G}} \nonumber \\
&& \delta(\hbar\omega_{\nu\mathbf{q}}+\hbar\omega_{\nu'\mathbf{q'}}-\hbar\omega_{\nu''\mathbf{q''}})
\end{eqnarray}
and
\begin{eqnarray}
    P_{\nu\mathbf{q}}^{\nu'\mathbf{q'},\nu''\mathbf{q''}} &=& \frac{2\pi}{\mathcal{N}_q \hbar^2} \sum_\mathbf{G} \left|V^{(3)}(\nu\mathbf{q},\nu'\mathbf{-q'},\nu''\mathbf{-q''}) \right|^2 \nonumber \\
&&    \times \bar{N}_{\nu\mathbf{q}} (\bar{N}_{\nu'\mathbf{q'}}+1) (\bar{N}_{\nu''\mathbf{q''}}+1) \delta_{\mathbf{q-q'-q'',G}} \nonumber \\
&& \delta(\hbar\omega_{\nu\mathbf{q}}-\hbar\omega_{\nu'\mathbf{q'}}-\hbar\omega_{\nu''\mathbf{q''}})
\end{eqnarray}
respectively represent the absorption or emission of the phonon mode $\nu'\mathbf{q'}$ when an initial phonon mode $\nu\mathbf{q}$ is scattered into $\nu''\mathbf{q''}$. Here $\mathcal{N}_\mathrm{q}$ is the number of $q$-points in a uniform mesh and $\mathbf{G}$ is a reciprocal lattice vector. Most significantly, $V^{(3)}$ is the appropriately defined Fourier transform of the third-order force constants in Eq.~(\ref{eq:3rd-order}) that can be determined using \emph{ab initio} methods. (See Ref.~\citenum{fugallo2013ab} for details.)

Higher-order derivatives of the total energy contain information about higher-order phonon-phonon processes. The inclusion of four-phonon scattering within the single mode relaxation time approximation (SMRTA) to the BTE was detailed in Ref.~\citenum{feng2016quantum}, and has been implemented in \texttt{FourPhonon} \cite{han2022fourphonon}, an extension to the \texttt{ShengBTE} package.
%Detailed expressions are given in Ref.~\citenum{cepellotti2022phoebe}.

\subsubsection{Isotope scattering}

The next term in the BTE, Eq.~(\ref{eq:iso}), describes scattering due to mass disorder, sometimes called isotope scattering. It is treated as scattering from an appropriately averaged point defect \cite{garg2011role}. 
\begin{eqnarray}
    P^\mathrm{isotope}_{\nu\mathbf{q},\nu'\mathbf{q'}} &=& \frac{\pi}{2\mathcal{N}_q} \omega_{\nu\mathbf{q}} \omega_{\nu'\mathbf{q'}} \left[ \bar{N}_{\nu\mathbf{q}} \bar{N}_{\nu'\mathbf{q'}} +\frac{1}{2} \left( \bar{N}_{\nu\mathbf{q}}+\bar{N}_{\nu'\mathbf{q}'}\right)\right] \nonumber \\
    && \times \sum_\kappa g^\kappa_2 \left| \sum_\alpha e^*_{\kappa\alpha,\nu}(\mathbf{q}) e_{\kappa\alpha,\nu'}(\mathbf{q'})\right|^2 \delta(\omega_{\nu\mathbf{q}} -\omega_{\nu'\mathbf{q'}})
\end{eqnarray}
The coupling strength $g^\kappa_2$ is an input, either an average over the natural isotopic mass distribution, or chosen by hand to simulate a particular doping scheme.

\subsubsection{Boundary scattering}

Boundary scattering due to the physical size of a crystal can be incorporated by a simple term that depends only on the size, the phonon group velocity, and the equilibrium populations: 
%Details are given in Ref.~\citenum{cepellotti2022phoebe}.
\begin{equation}
P^\mathrm{boundary}_{\nu\mathbf{q}} =   \frac{\mathbf{v}_{\nu\mathbf{q}}}{LF} \bar{N}_{\nu\mathbf{q}} (\bar{N}_{\nu\mathbf{q}}+1).
\end{equation}
Here $L$ is the Casimir length and $F$ is a geometric correction based on the aspect ratio of the sample.\cite{fugallo2013ab} 

\subsubsection{Phonon-electron scattering}

The same electron-phonon coupling that leads to scattering of electrons by phonons can also contribute to scattering of phonons by electrons, given by the last term in the linearized BTE.\cite{cepellotti2022phoebe} The phonon-electron scattering rate is determined by the imaginary part of the phonon self-energy diagram  that contains a virtual electron-hole pair, shown in Fig.~\ref{fig:diagrams}(e). 
\begin{equation}
    P^\mathrm{ph-el}_{\nu\mathbf{q}} = -\frac{2\pi}{\mathcal{N}_k \hbar} \bar{N}_{\nu\mathbf{q}} (\bar{N}_{\nu\mathbf{q}}+1) \sum_{mn\mathbf{k}} \left| g_{mn\nu}(\mathbf{k},\mathbf{q}) \right|^2 (f_{n\mathbf{k}}-f_{m\mathbf{k+q}}) \delta(\epsilon_{n\mathbf{k}} - \epsilon_{m\mathbf{k+q}} -\omega_{\nu\mathbf{q}}),
\end{equation}
where $g_{mn\nu}(\mathbf{k},\mathbf{q})$ is the electron-phonon coupling defined in Eq.~(\ref{gkq}).
These contributions are often assumed to be smaller than the phonon-phonon scattering, but they have been shown to make a significant contribution in some metals or highly doped semiconductors \cite{liao2015significant}. They are not included in the examples discussed below.

\subsection{Combining scattering processes through Matthiessen's rule}

Within the RTA, if there are several scattering mechanisms that are approximately independent, their respective scattering times can be combined using Matthiessen's rule. For example, for charge carriers subject to non-polar and polar phonon scattering, as well as scattering from charged impurities, the total scattering rate that appears in the BTE, Eq.~(\ref{coll}) or (\ref{boltz3}), will be:
\begin{equation}
\label{Mathiessen}
\frac{1}{\tau_\mathrm{tot}} = \frac{1}{\tau_\mathrm{npol}}
+ \frac{1}{\tau_\mathrm{pol}} + \frac{1}{\tau_\mathrm{imp}}~.
\end{equation}
The temperature dependence of the RT is given indirectly through the phonon and electron distributions 
within Eq.~(\ref{imag}). Additionally, for $\tau_\mathrm{pol}$ and $\tau_\mathrm{imp}$, $T$ and $\mu$ dependence enters 
implicitly through their respective screening radii ($r_{\infty}$ and $r_0$) as defined in Eq.~(\ref{r0inf}). This dependence on $\mu$ allows for the study of
doped materials, which are important for the optimization of $zT$ for thermoelectric applications.

\section{Numerical approaches and post-processing}
\label{sec:numerics}

In practice, the calculations of TE transport properties on the basis of RTA-BTE can be performed using different levels of approximation with increasing computational cost. The constant relaxation time approximation (CRTA)\cite{madsen2006boltztrap} or methods based on the deformation potential approximation (DPA),\cite{bardeen1950deformation} have been used 
extensively.\cite{ xi2012first,xi2018discovery,ganose2021efficient,chaves2021investigating} The CRTA is generally a poor choice to describe properties other than the Seebeck coefficient due to the lack of el-ph information, and DPA methods fail drastically for materials presenting strong polar optical phonon scattering or inter-band scatterings. 
Thus, full first-principles calculations of el-ph coupling is necessary in order to obtain accurate TE transport properties.  

Even after the sum over phonon modes and integration over the phonon BZ, the el-ph RTs in Eq.~(\ref{imag}) are a rich source of microscopic information since they are resolved for different bands and ${\bf{k}}$-points.
However, calculation of macroscopic transport properties requires a further sum over electron bands and integration over the electron BZ. 
In order to converge the double BZ integration, very dense ${\bf{k}}$ and ${\bf{q}}$ meshes are needed, since the denominator of the integrand may exhibit significant fluctuations on the scale of the phonon energy.
For example, calculated mobilities for Si required 120$\times$120$\times$120 ${\bf{k}}$/${\bf{q}}$ final meshes in order to reach a convergence criterion of $10^{-4}$ \cite{ma2018first}.
Though calculations on such dense meshes have recently been done directly with DFPT\cite{brunin2020phonon}, they remain computationally demanding, and different interpolation schemes are frequently used\cite{giustino2017electron,giustino2007electron,li2015electrical,Agapito2018}.
Here we briefly review the Wannier-Fourier (WF) and Dual Interpolation approaches. 

There are alternative approaches that attempt to reduce the computational
cost of el-ph calculations: Samsonidze and Kozinsky proposed the el–ph averaged
(EPA) approximation to be used mostly for isotropic materials\cite{samsonidze2018accelerated}, while Deng {\it{et al.}} proposed an approach using a generalized Eliashberg function for short-range el-ph coupling and analytical expressions for long-range el–ph and e–d scatterings\cite{deng2020epic}. 
Such methods, along with the development of efficient interpolation schemes, facilitate automated and
unsupervised predictions of novel functional materials for thermoelectric or other electronic applications via high-throughput screening.\cite{xi2018discovery,samsonidze2018accelerated,deng2020epic,yao2021materials}

\subsection{Interpolation schemes}\label{sec:interpolation}
        
WF interpolation was introduced by Giustino, Cohen and Louie~\cite{giustino2007electron} and is based on maximally localized Wannier functions (MLWF) \cite{marzari1997maximally} and is analogous to the strategy developed for getting phonon dispersion relations using the interatomic force constants \cite{gonze1997dynamical}.
Within this approach, electronic band structure, phononic dispersions, and el-ph matrix elements, $g(\mathbf{k},\mathbf{q})$, calculated using DFT and DFPT on coarse $\mathbf{k},\mathbf{q}$
grids, are interpolated onto much finer $\mathbf{k}',\mathbf{q}'$
grids through simple matrix multiplications \cite{giustino2007electron}.
The el-ph matrix elements on the fine grid are given by
\begin{equation} 
g({\bf{k^ {\prime}}},{\bf{q^ {\prime}}}) = \frac{1}{N_e} \sum_{{\bf{R}}_e,{\bf{R}}_p} e^{i({\bf{k^{\prime}}}\cdot{\bf{R}}_e+{\bf{q^{\prime}}}\cdot{\bf{R}}_p)}{\bf{U}}_{{\bf{k^{\prime}}}+{\bf{q^{\prime}}}}g({\bf{R}}_e,{\bf{R}}_p){\bf{U}}_{{\bf{k^{\prime}}}}^{\dagger} {\bf{u}}_{\bf{q^{\prime}}},
\label{g_interpolation}
\end{equation}
where ${\bf{R}}_{e}$ and ${\bf{R}}_{p}$ are primitive lattice vectors
of the Wigner-Seitz (WS) supercell with Born-von-K{\'a}rm{\'a}n periodic boundary conditions and
{\bf{U}}$_{{\bf{k^{\prime}}}}$ ({\bf{u}}$_{{\bf{q^{\prime}}}}$) are diagonalizer matrices
over ${\bf{k^{\prime}}}$ (${\bf{q^{\prime}}}$) indices from Wannier to Bloch representations
for electrons (phonons). The el-ph matrix elements in the real-space Wannier representation are
\begin{equation}
g({\bf{R}}_e,{\bf{R}}_p) = \frac{1}{N_p} \sum_{{\bf{k}},{\bf{q}}} e^{-i({\bf{k}}\cdot{\bf{R}}_e+{\bf{q}}\cdot{\bf{R}}_p)}{\bf{U}}_{{\bf{k}}+{\bf{q}}}^{\dagger}g({\bf{k}},{\bf{q}}){\bf{U}}_{{\bf{k}}} {\bf{u}}_{{\bf{q}}}^ {-1}~,
\end{equation}
{\bf{U}}$_{{\bf{k}}}$ is an unitary matrix corresponding to
the rotation of the corresponding electronic states from
Bloch to Wannier representations within the gauge of
MLWF and {\bf{u}}$_{{\bf{q}}}$ is a phonon eigenvector.
In the above equations, electron band and phonon branch indices
are omitted for simplicity.
WF interpolation is variously implemented in the \texttt{EPW} \cite{ponce2016epw}, \texttt{VASP} \cite{engel2020electron}, and \texttt{Perturbo} \cite{zhou2021perturbo} codes and is the foundation upon which Dual Interpolation was developed. 

The accuracy of WF interpolation strongly depends
on the spatial localization of $g({\bf{R}}_e,{\bf{R}}_p)$, which makes it possible to neglect matrix elements outside
the WS supercell generated from the initial coarse BZ mesh.
A more detailed analysis suggests $g({\bf{R}}_e,{\bf{R}}_p)$should decay in the variable ${\bf{R}}_e$ at least as fast as MLWFs. In fact, MLWFs in insulators decay quickly provided 2D and 3D systems present time-reversal symmetry~\cite{brouder2007exponential}. 
For metals, localized Wannier functions can be obtained from the disentanglement procedure \cite{souza2001maximally}.
For ${\bf{R}}_e = 0$ the localization depends strongly on the dielectric properties of the material and $g(0,{\bf{R}}_p)$ decays with ${\bf{R}}_p$ due to the screened Coulomb interaction
of the potential generated by
atomic displacement. 
In particular, metals present short screening lengths based on Friedel oscillations and decay as $|{\bf{R}}_p|^{-4}$\cite{fetter2012quantum}, while nonpolar semiconductors may possess an incomplete screening and decay at the rate of a quadrupole, $|{\bf{R}}_p|^{-3}$~\cite{pick1970microscopic}.
For ionic and polar covalent crystals, the interpolation turns to be more intricate since 
the long-range Fr{\"{o}}hlich dipole coupling is a relevant contribution,
which decays as $|{\bf{R}}_p|^{-2}$, resulting a well-known $|{\bf{q}}^{-1}|$ divergence in momentum space when $|{\bf{q}}|\rightarrow0$ \cite{vogl1976microscopic}.
%The WF interpolation scheme developed to treat dipole and also quadrupole interactions will be discussed in the next section.
The method for treating dipole and quadrupole interactions within the WF scheme was outlined in Sect.~\ref{sec:el-ph}.

\subsection{\texttt{turboEPW} with dual interpolation}
       
The dual interpolation approach was developed recently by several of the present authors~\cite{chaves2020boosting} and represents an effort to improve the computational performance of WF interpolation. It is based on two sequential interpolations, namely, WF interpolation followed by a Fourier interpolation based on star functions. The implementation, called \texttt{Turbo-EPW},  was built as an extension of the \texttt{EPW} code \cite{ponce2016epw} in order to make use of the latter's well-tested WF interpolation.

As mentioned previously, the calculation of transport properties requires a double integration over ${\bf{k}}$ and ${\bf{q}}$ wave vectors. The idea of dual interpolation is to determine $g$ over a
fine $\mathbf{q}'$ grid using WF interpolation, and perform the partial integration at
each of the $n_\mathbf{\bar{k}}$ \emph{irreducible} k-points, ${\bf{\bar{k}}}_l$,
corresponding to a moderately sized, regular k-mesh (${\bf{k^r}}$). (For reasons of clarity, we start with $\mathbf{q}'$ integration first, however one can easily switch the order and start with $\mathbf{k}'$ integration.) 
In this way we determine a generic transport function, $f({\bf{\bar{k}}}_l)$, already integrated over a fine $\mathbf{q}'$ grid. The next step is the calculation of $f$ over the whole BZ with fine $\mathbf{k}'$ grid using a suitable second interpolation that needs to take into account the symmetry of the crystal. 
The second interpolation uses symmetry-adapted plane-waves or star functions, $\Upsilon_m({\bf{k^{\prime}}})$,
as a basis set to Fourier expand $f$~\cite{chadi1973special}:
\begin{equation} \label{expansion}
\tilde{f}({\bf{k^ {\prime}}}) = \sum_{m=1}^M a_m \Upsilon_m({\bf{k^{\prime}}})~,
\end{equation}
where
$\Upsilon_m({\bf{k^ {\prime}}})=\frac{1}{n_s}\sum_{\{\upsilon\}}\exp[{i(\upsilon {\bf{R}}_m)\cdot{\bf{k^{\prime}}}}]$,
with the sum running over all $n_s$ point group symmetry operations $\{\upsilon\}$
of the direct lattice ${\bf{R}}_m$.

Following the method first proposed by Shankland-Koelling-Wood \cite{shankland1971interpolation,koelling1986interpolation},
the number of star functions in the expansion, $M$,
is taken to be greater than the number of data points ($M > n_{{\bf{\bar{k}}}}$), and the interpolating function, $\tilde{f}$, is required to pass through
the data points exactly. The freedom from
extra star functions is used to minimize a
spline-like roughness functional in order to minimize oscillations between data points.
As defined by Pickett, Krakauer and Allen~\cite{pickett1988smooth}, the spline-like roughness functional reads
$\label{R}
\Pi = \sum_{m=2}^M \lvert a_m \rvert ^2 \rho(R_m)$
with
$\rho(R_m) = \left(1-c_1\left({\frac{R_m}{R_{min}}}\right)^2\right)^2+c_2(\frac{R_m}{R_{min}})^6~,$
where $R_m = \lvert {\bf{R}}_m \rvert$, $R_{min}$ is the magnitude
of the smallest nonzero lattice vector, and $c_1 = c_2 = 3/4$.
The determination of the Fourier coefficients, $a_m$, is accomplished by the Lagrange multiplier method once the problem has been reduced to minimizing ${\Pi}$ subject to the constraints, $\tilde{f}({\bf{\bar{k}}}_l)={f}({\bf{\bar{k}}}_l)$.
The result is
\begin{equation}
a_m = \begin{cases}
\rho(R_m)^{-1} \sum_{l=1}^{n_{\bf{\bar{k}}}-1} \lambda^*_l\left[\Upsilon_m^{*}({\bf{\bar{k}}}_l) - \Upsilon_m^{*}({\bf{\bar{k}}}_{n_{{\bf{\bar{k}}}}})\right],  &  m>1, \\
f({\bf{\bar{k}}}_{n_{\bf{\bar{k}}}})-\sum_{m=2}^M a_m \Upsilon_m ({\bf{\bar{k}}}_{n_{{\bf{\bar{k}}}}}),  &  m=1,
\end{cases}
\end{equation}
in which the Lagrange multipliers, $\lambda^*_l$, can be evaluated from
\begin{equation}
f({\bf{\bar{k}}}_p) - f({\bf{\bar{k}}}_{n_{{\bf{\bar{k}}}}}) = \sum_{l=1}^{n_{{\bf{\bar{k}}}}-1}{\bf{H}}_{pl}\lambda^*_l~,
\end{equation}
with
\begin{equation}
{\bf{H}}_{pl} = \sum_{m=2}^M \frac{\left[\Upsilon_m({\bf{\bar{k}}}_p) - \Upsilon_m({\bf{\bar{k}}}_{n_{{\bf{\bar{k}}}}})\right]\left[\Upsilon_m^{*}({\bf{\bar{k}}}_l)-\Upsilon_m^{*}({\bf{\bar{k}}}_{n_{{\bf{\bar{k}}}}})\right]}{\rho(R_m)}~.
\end{equation}

The interpolating function $\tilde{f}$ can be written as a linear mapping of the WF data
\begin{equation}
\tilde{f}({\bf{k^{\prime}}}) = \sum_{l=1}^{n_{{\bf{\bar{k}}}}-1}J({\bf{\bar{k}}}_l,{\bf{k^{\prime}}}) [f({\bf{\bar{k}}}_l) - f({\bf{\bar{k}}}_{n_{{\bf{\bar{k}}}}})]~,
\end{equation}
where $J$ is the transformation formula 
independent of the data. In fact, $J$ is
determined by the set of irreducible
sampling points (${\bf{\bar{k}}}_l$), the number of star functions ($M$), and
the form of the roughness functional ($\Pi$):
\begin{equation}
J({\bf{\bar{k}}}_l,{\bf{k^{\prime}}}) = \sum_{p=1}^{n_{{\bf{\bar{k}}}}-1}\sum_{m}^{M}\frac{[\Upsilon_{m}^{*}({\bf{\bar{k}}}_p)-\Upsilon_{m}^{*}({\bf{\bar{k}}}_{n_{{\bf{\bar{k}}}}})]\Upsilon_{m}({\bf{k^{\prime}}})}{\rho(R_m){\bf{H}}_{pl}}~.
\end{equation}
Basically, $J$ transforms ${\bf{\bar{k}}}_l \rightarrow {\bf{k^{\prime}}}$, which allows for great computational savings since the final homogeneous grid (${\bf{k^{\prime}}}$) on which $f$ is calculated 
can be much larger than the regular grid (${\bf{k^{r}}}$) that generated the irreducible points.

In practice, to expand the interpolating function in Eq.~(\ref{expansion}), we rely on a 3D Fast Fourier Transform (FFT) to reciprocal space of the lattice points and their respective star functions that were generated in real space.
We take advantage of the periodic boundary conditions to enlarge the real space by the expansion factor $M$, the number of star functions per ${\bf{k}}$-point, to get  a new homogeneous ${\bf{k^{\prime}}}$-grid much finer than the original one.
To take into account crystal anisotropy, the extension of the real space is determined by defining spheres for each crystallographic axis with the maximum radius given in terms of their reciprocal primitive vectors. More details can be found in Ref.~\citenum{chaves2020boosting}.

The FFT computational complexity, $\mathcal{O}(N\log{}N)$, where $N$ corresponds to the number of data points related to the product of FFT dimensions, is more affordable than the computational complexity of classical matrix multiplications ($\mathcal{O}(N^3)$) as performed by single WF interpolation. 
The overall gain in  computational performance by using the dual interpolation method compared to a single WF interpolation is approximately $2 (n_s \times M)$. 
As $M$ typically ranges between $5$ and $60$, there is a great boost in performance that allows for improved calculations of el-ph mediated transport properties. 
This method was employed to calculate thermoelectric properties in layered materials, such as SnSe and GeSe and will be reviewed in Sect.~\ref{sec:application}.  

\subsection{Third order force constants}

The calculation of lattice thermal conductivity within the BTE framework requires determination of the relevant contributions to the phonon scattering matrix. The primary contribution is the phonon-phonon scattering based on the calculation of third-order force constants. 
While in principle these could be calculated using DFPT, in practice they are generally determined using the supercell method. Routines within either the \texttt{ShengBTE} \cite{wu2014shengbte} or \texttt{phono3py} \cite{phono3py} packages can be used create the necessary supercells containing strategically displaced atoms that allow for the construction of the third-order force constants after the the total energy DFT calculations for each of the supercells has been completed. 
Since the number of atomic triplets grows rapidly, a cutoff distance, beyond which atomic interactions are ignored, is generally applied to keep the number of supercell calculations manageable. 
However, for materials with low thermal conductivity, care must be taken to converge results with respect to the cutoff, since long-range interactions can be important for accurate determinations of the lattice thermal conductivity \cite{carrete2014low}.

\subsection{Calculating transport coefficients}
\label{sec:coefficients}

Once the ingredients for the electron/hole or phonon scattering matrix have been determined, the BTE needs to be solved to find the transport coefficients and properties of interest. Within the RTA the \texttt{BoltzTraP} package\cite{madsen2006boltztrap} is frequently employed for determining electron and hole transport properties, while the \texttt{ShengBTE} package \cite{wu2014shengbte} is often used for phonons to calculate thermal transport. The \texttt{Perturbo} \cite{zhou2021perturbo} and \texttt{Phoebe} \cite{cepellotti2022phoebe} codes seek to treat both thermal and electrical conduction within a single framework. 
    
\section{Application to layered thermoelectric materials}
\label{sec:application}

\subsection{Thermoelectric $zT$ optimization}

 Thermoelectric performance, as measured by the dimensionless figure of merit $zT = \sigma S^2 T / \kappa_\mathrm{tot}$, is maximized at a given temperature $T$ in materials with both a high electrical conductivity $\sigma$ and high Seebeck coefficient $S$, along with low total thermal conductivity, $\kappa_\mathrm{tot} = \kappa_\mathrm{carr} + \kappa_\mathrm{latt}$, which has contributions from the electrical carriers and the atomic lattice, respectively.
Because the calculation of the thermoelectric figure of merit depends on both electronic and phononic transport properties, it serves as an excellent example of how the above methodologies can be put into practice. 

%A thermoelectric with a high figure of merit is expected to be a material with a high $PF$, where the charge-coupled entropy conductivity is larger than the non-coupling entropy conductivity due to purely irreversible processes. 
One strategy to increase $zT$ is to maximize the $PF$ through band-structure 
engineering.\cite{pei2011convergence,pei2012band,liu2012convergence,dehkordi2015thermoelectric,parker2015benefits}
On the other hand, in order to decrease purely irreversible processes, common strategies focus on the reduction of the lattice thermal conductivity, $\kappa_\mathrm{latt}, $\cite{morelli2008intrinsically,he2016ultralow,gonzalez2018ultralow,hochbaum2008enhanced,boukai2008silicon,kanatzidis2009nanostructured,zhao2013high} with less attention given to $\kappa_\mathrm{el}$. Given the high carrier concentrations of optimally doped TE materials, $\kappa_\mathrm{el}$ should not be ignored; reducing it
can be best accomplished by 
minimizing the Lorenz number, $L = \kappa_\mathrm{el}/(\sigma T)$.\cite{mckinney2017search,mahan1996best} 
It is important to note, however, that direct measurements of $L$ and $\kappa_\mathrm{el}$ are nontrivial. Typically $\kappa_\mathrm{el}$ is estimated based on the Wiedemann–Franz law by using
measured values of $\sigma$ and estimated values of $L$ from simplified parabolic band approximations.\cite{baranowski2013effective,ortiz2017potential} 
In general such estimations are inaccurate,\cite{mckinney2017search,putatunda2019lorenz} making first-principles calculations of $L$ and $\kappa_\mathrm{el}$ necessary.  

\begin{figure}[tbp]
\includegraphics[width=0.8\columnwidth]{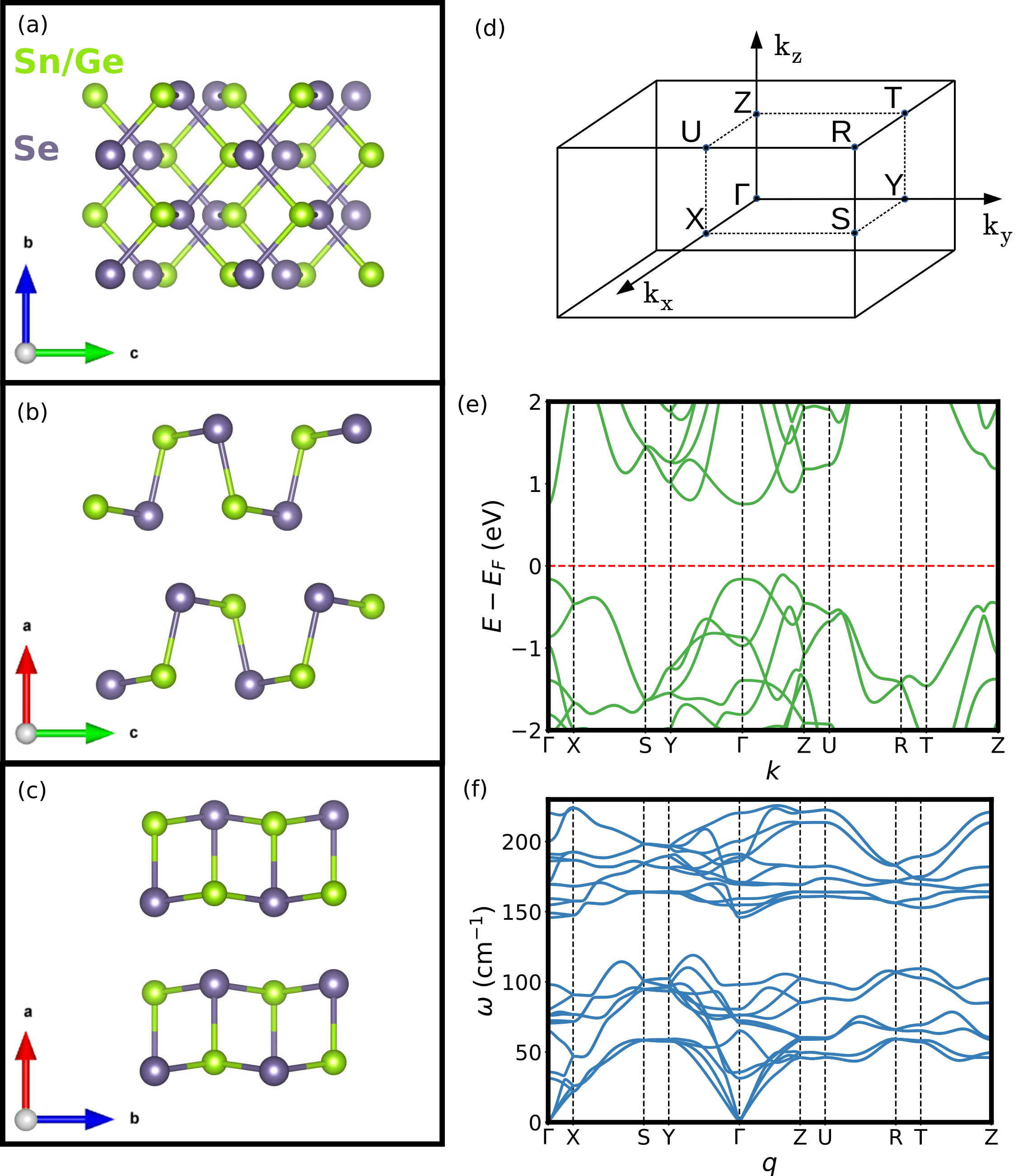}
\caption{\label{fig:GeSe} (a)-(c) GeSe (SnSe) crystal structure along each axis. (d) The Brillouin zone with high symmetry points labeled. (e) Electronic and (f) phononic band structure for GeSe.}
\end{figure}

Despite the complexity arising from the interdependence of 
all the transport properties that contribute to $zT$, 
the search for high-$zT$ materials continues \cite{he2017advances},
and new high-performance TE materials are 
constantly emerging.\cite{biswas2012high,liu2012copper,fu2016enhanced,olvera2017partial,cheng2017new,ma2020alpha,roychowdhury2021enhanced} An important class of such materials are the bulk crystals 
with a two-dimensional (2D) layered structure which have high anisotropy and 
improved electrical conductivity along in-plane directions.\cite{terasaki1997large,rhyee2009peierls,ohta2018high,cheng2019optimal} (For a review see Li {\it{et al.}}\cite{li2022layered})
In particular, the extremely high $zT$ values reported for intrinsic~\cite{zhao2014ultralow}, p-doped \cite{zhao2016ultrahigh} and n-doped\cite{chang20183d} SnSe has boosted the interest in high-efficiency layered TE materials.
Significant effort has gone into trying to accurately calculate its properties \cite{carrete2014low,ding2015high,guo2015first,skelton2016anharmonicity,li2019resolving,aseginolaza2019phonon,chaves2021microscopic}, both in order to optimize its performance, but also in the hopes that understanding its fundamental properties will allow for predictions of additional high-performance thermoelectric materials. In that vein, germanium selenide (GeSe) is an obvious isostructural material that has been the focus of only a handful of experimental \cite{zhang2016thermoelectric,shaabani2017thermoelectric} and theoretical \cite{hao2016computational,roychowdhury2018germanium,yuan2019tailoring,chaves2022out} studies. 
The crystal structure of GeSe (SnSe) is shown in Fig.~\ref{fig:GeSe}, along with the basic electronic and phononic band structures calculated using DFT.
Here we discuss calculations of the thermoelectric properties of SnSe and GeSe in comparison with previous calculations and experimental measurements in order to highlight the successes and challenges inherent in a DFT-based first-principles framework for calculating thermoelectric performance.

\subsection{Electronic transport in SnSe and GeSe}      

\begin{figure}[htbp]
\includegraphics[width=\columnwidth]
{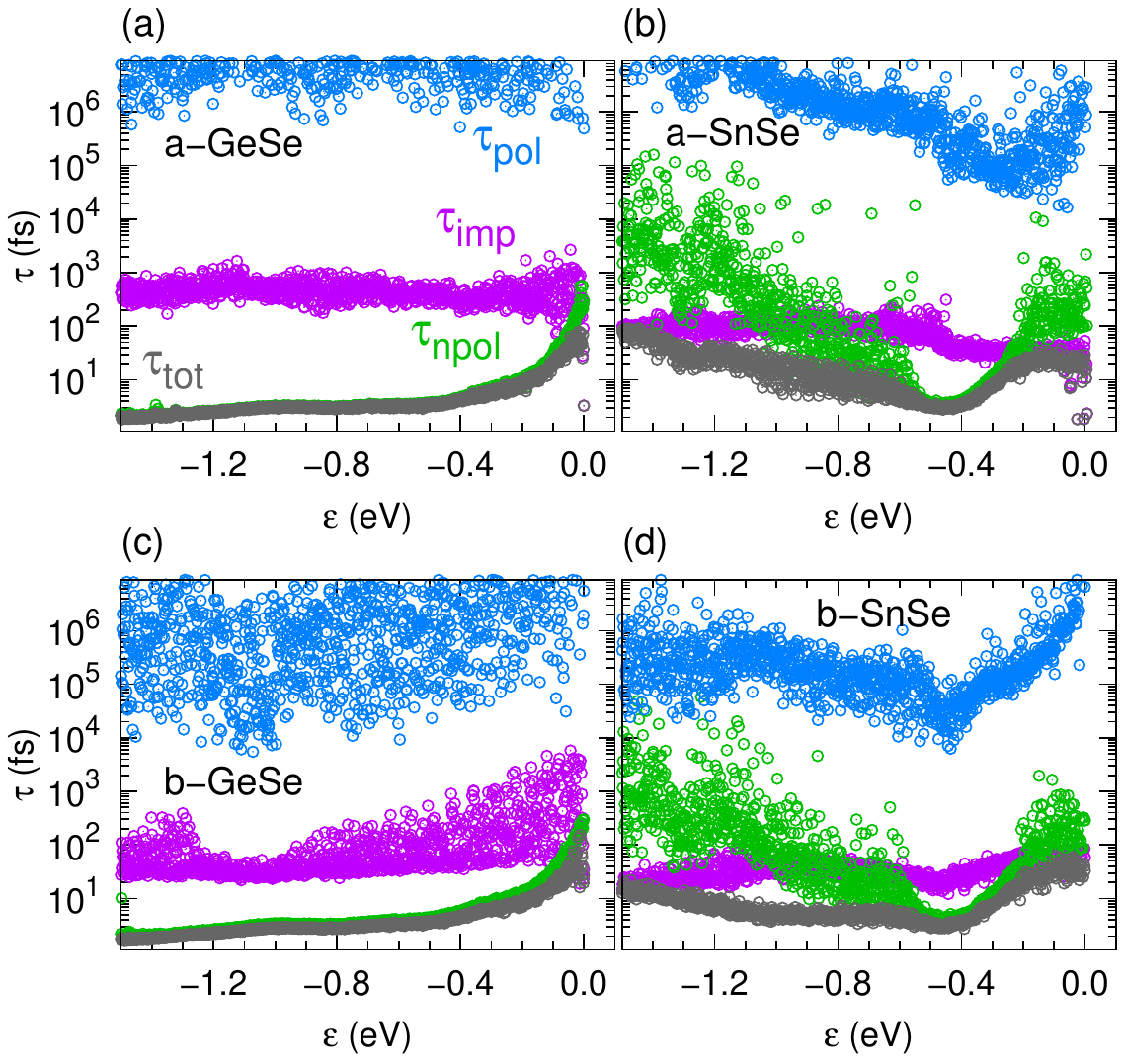}
\caption{\label{fig:tau} Energy-resolved scattering times due to polar and non-polar phonons as well as impurities, calculated for the a- and b-axes of SnSe and GeSe. From Ref.~\citenum{chaves2022out}.}
\end{figure}

Using the above framework for calculating electron-phonon scattering with DFT, our \texttt{TurboEPW} implementation \cite{chaves2020boosting} allowed for the sampling of over 1 billion $\mathbf{k/q}$ pairs and thus a detailed calculation of the momentum- and band-resolved impurity, polar, and non-polar scattering rates. These scattering rates can be expressed as a function of the carrier energy using a velocity-weighted average:
\begin{equation}
    \tau(\epsilon) = \frac{\sum_{n,\mathbf{k}} \tau(n,\mathbf{k}) v_{n,\mathbf{k}} v_{n,\mathbf{k}} \delta(\epsilon - \epsilon_{n,\mathbf{k}})} {\sum_{n,\mathbf{k}} v_{n,\mathbf{k}} v_{n,\mathbf{k}} \delta(\epsilon - \epsilon_{n,\mathbf{k}})},
\end{equation}
where $v_{n,\mathbf{k}} = \partial \epsilon_{n,\mathbf{k}}/\partial{k}$ is the carrier velocity. Scattering times as a function of carrier energy for SnSe and GeSe are shown in Fig.~\ref{fig:tau}, with more detailed discussion provided in Ref.~\citenum{chaves2022out}. The BTE was solved within RTA-SERTA approximation.

\begin{figure}[htbp]
\includegraphics[width=0.60\columnwidth]
{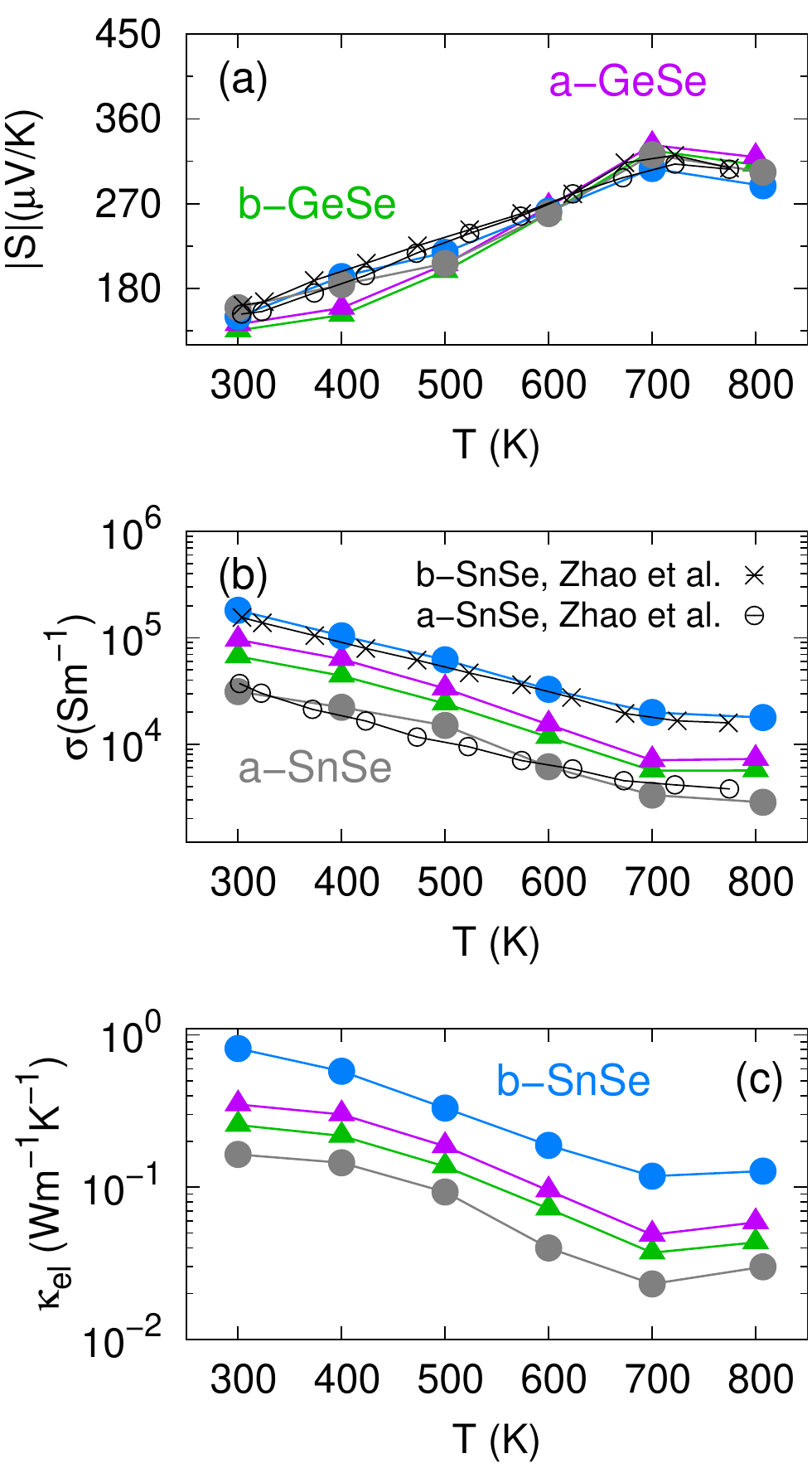}
\caption{\label{fig:properties} Thermoelectric transport properties for the a- and b-axes of SnSe and GeSe as a function of temperature, with hole doping concentrations based on experiment.}
\end{figure}

The \texttt{BoltzTraP} package \cite{madsen2006boltztrap} uses the energy-resolved scattering times to determine many transport properties, including the Seebeck coefficient $S$, the electrical conductivity $\sigma$, and the carrier thermal conductivity $\kappa_\mathrm{carr}$, which are needed for preditions of the thermoelectric figure of merit, $zT$. Fig.~\ref{fig:properties} shows these properties as a function of temperature for the $a$- and $b$-axes of SnSe and GeSe with hole doping concentrations based on experimental measurements.\cite{chaves2022out}. 

\subsection{Lattice thermal transport in SnSe and GeSe}

The extremely low intrinsic lattice thermal conductivity in single crystal SnSe is a significant contributor to its high figure of merit, so it is natural to hope that GeSe might also exhibit a similarly low $\kappa_\mathrm{latt}$. The Debye-Callaway framework \cite{asen1997thermal,hao2016computational} allows for a simple estimate of $\kappa_\mathrm{latt}$ based on calculations of the second-order force constants needed to determine the phonon spectrum of a crystal. The acoustic phonon velocities and Gr\"uneisen parameters (volume dependence of the phonon frequencies) are used in conjunction with a model for the normal and umklapp scattering rates of acoustic phonons. However, the simplest fully first-principles calculation of $\kappa_\mathrm{latt}$ uses third-order force constants calculated within the harmonic approximation in order to determine the detailed band- and momentum-resolved three-phonon contribution to the scattering matrix for use in the phonon BTE. Additional terms for phonon-isotope, phonon-boundary, and phonon-electron scattering can also be taken into account in the scattering matrix.

Here we present a new calculation of the lattice thermal conductivities of SnSe and GeSe using an iterative solution to the Boltzman Transport Equation (BTE) \cite{omini1995iterative}
including third-order force constants for very distant neighbors in large supercells. We include corrections based on the Wigner distribution \cite{simoncelli2019unified}, which gives small but non-negligible corrections arising from phonon bands that overlap due to their finite linewidths. 

\begin{figure}[tbp]
\includegraphics[width=0.49\columnwidth]{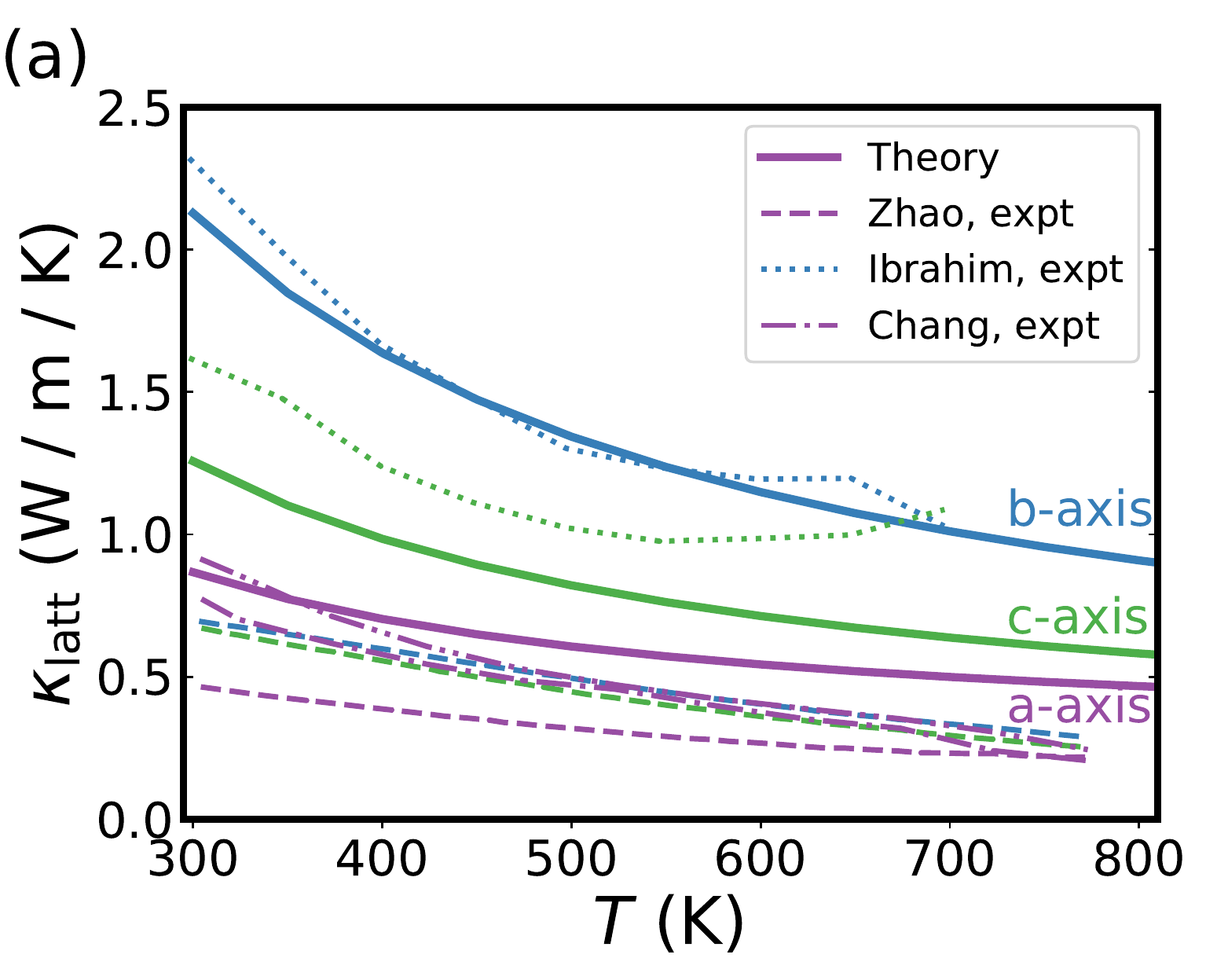}
\includegraphics[width=0.49\columnwidth]{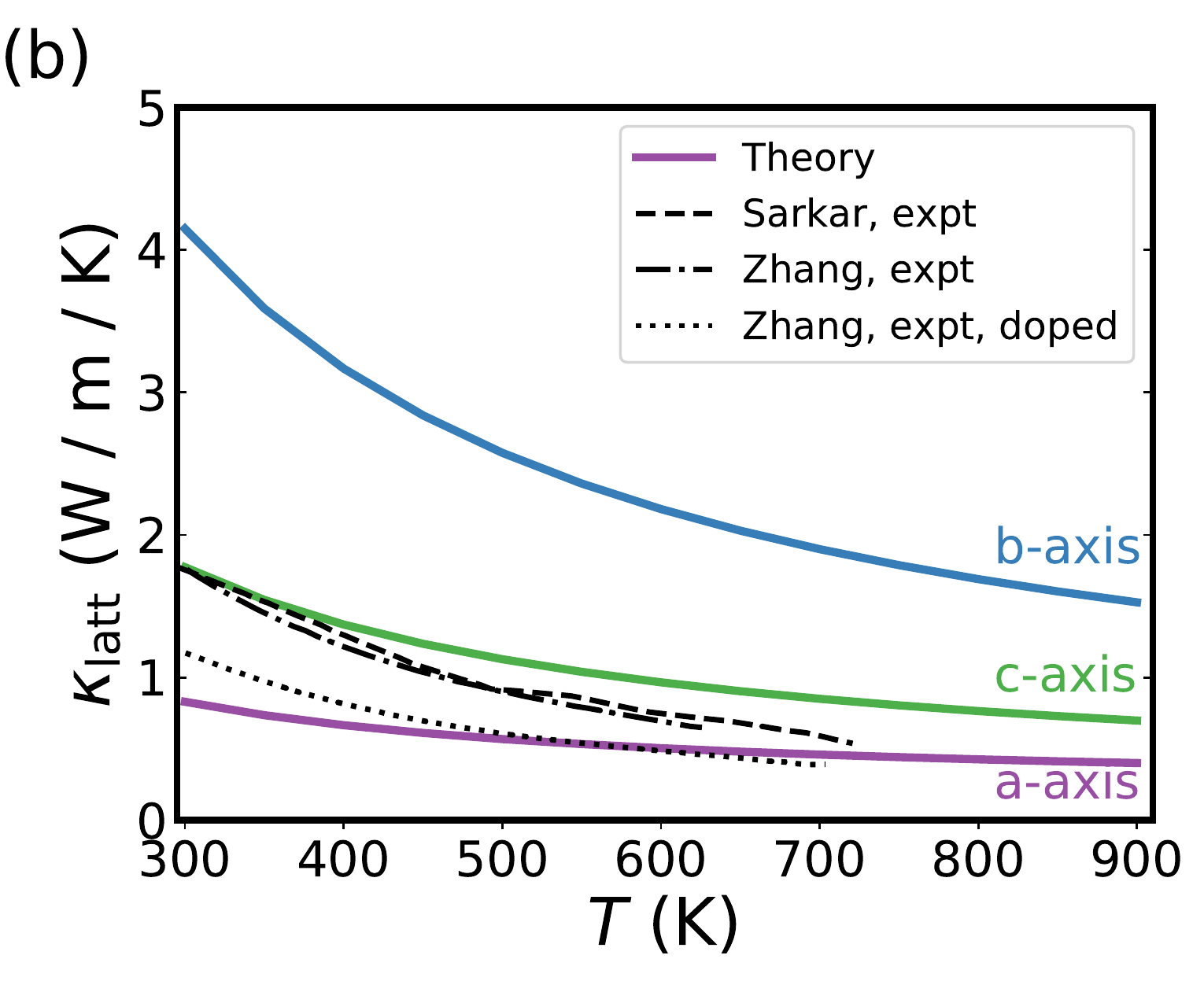}
\caption{\label{fig:kappa-vs-expt} Calculated lattice thermal conductivity along each axis as a function of temperature, compared to published experimental measurements for (a) SnSe and (b) GeSe. Experimental SnSe measurements are from Zhao \cite{zhao2016ultrahigh}, Ibrahim \cite{ibrahim2017reinvestigation}, and Chang \cite{chang20183d} while GeSe measurements are from from Sarkar \cite{sarkar2020ferroelectric} (polycrystalline) and Zhang \cite{zhang2016thermoelectric} (polycrystalline GeSe without and with 3\% Ag doping).}
\end{figure}

We begin with the results for SnSe because there is a large body of literature seeking to reconcile different experimental and theoretical results, yielding many sources for comparison. 
As shown in Fig.~\ref{fig:GeSe}(a)-(c), the SnSe crystal is highly anisotropic, so the results for $\kappa_\mathrm{latt}$ are calculated separately along the a-, b-, and c-axes and displayed in purple, blue, and green, respectively, in Fig.~\ref{fig:kappa-vs-expt}(a).
The force constants were calculated using \texttt{VASP}\cite{kresse1996efficient,kresse1996efficiency} with supercells generated by \texttt{phono3py}\cite{phono3py}, while the solution of the BTE and calculation of $\kappa_\mathrm{latt}$ was done with \texttt{Phoebe}\cite{cepellotti2022phoebe}.

Generally speaking, calculations predict a very low thermal conductivity for SnSe, but not quite as low as the initial measurements on single crystals \cite{zhao2014ultralow}, which were surprising because they were even lower than previous measurements on polycrystalline samples, which one would expect to have lower thermal conductivity due to increased boundary scattering. Subsequent work has suggested that careful removal of SnO residue from polycrystalline samples can reduce its thermal conductivity \cite{zhou2021polycrystalline}. In Fig.~\ref{fig:kappa-vs-expt}(a) our predictions are compared to several experimental measurements of both undoped and doped samples. While there is obviously not perfect agreement (even between experimental measurements), it is clear that the calculations are giving a reasonable estimate for the extremely low thermal conductivity of SnSe, especially if considered in the context of other insulators that have thermal conductivities that are orders of magnitude larger. Our calculated values for $\kappa_\mathrm{latt}$ are qualitatively similar though slightly larger than previous calculations in the literature that also use third-order force constants to study SnSe.\cite{carrete2014low,skelton2016anharmonicity}

Turning now to GeSe, Fig.~\ref{fig:kappa-vs-expt}(b) shows our calculated results for $\kappa_\mathrm{latt}$ in comparison to several experimental results. 
Previous calculations based on the Debye-Callaway theory
predict extremely low values of $\kappa_\mathrm{latt}$, well below 1 W/m/K for all three axes \cite{hao2016computational}.
Our results, based on the methodology described above,
yield values significantly higher and qualitatively different from those based on Debye-Callaway theory, but similar to previous calculations also based on third-order force constants \cite{yuan2019tailoring}, but without the inclusion of the Wigner correction. 
In particular, a very high level of anisotropy is predicted, with the $b$-axis $\kappa_\mathrm{latt}$ roughly double the value for $c$-axis, which is in turn roughly double the $a$-axis value, all at 300~K.

Calculations of lattice thermal conductivity, particularly for systems with significant anharmonicity, remain quite challenging. Within a specific framework it is essential to thoroughly converge the results with respect to the various computational parameters, such as supercell size, cutoff radius, DFT settings, and q-grid for the BTE, to list some examples relevant to above examples. However, different levels of theory can lead to significant variations in the results, without an \emph{a priori} means of gauging accuracy with respect to experiment. For instance, in a study of PbTe \cite{xia2018revisiting} the authors found that finite temperature phonon frequency shifts increased the lattice thermal conductivity, but the addition of 4-phonon scattering reduced the thermal conductivity. This is a case where accidental cancellation between higher-order contributions allowed simpler models to fortuitously agree well with experiment. 
The calculations and even experimental measurements of $\kappa_\mathrm{latt}$ for SnSe are also not simple and not without controversy \cite{wei2016intrinsic,zhao2016intrinsic,wu2017direct,ibrahim2017reinvestigation}, and significant effort has been put into increasing the sophistication of theoretical calculations by including, among other approaches, non-perturbative anharmonic effects \cite{aseginolaza2019phonon}.

Nevertheless, comparisons between different materials at the same level of theory can still give some important physical insight. 
Comparing the two panels of Fig.~\ref{fig:kappa-vs-expt} we see that while the thermal conductivity along the $a$-axis is nearly identical for SnSe and GeSe, the latter exhibits slightly higher thermal conductivities along the $b$- and $c$-axis. 
Averaging over the three axes, as would be relevant for polycrystalline samples, the GeSe thermal conductivity is roughly 50\% larger than SnSe throughout the temperature range studied. 
There has been recent success purifying polycrystalline SnSe to remove tin oxides \cite{lee2019surface,zhou2021polycrystalline}, revealing the intrinsic lattice thermal conductivity that matches more closely the experimental measurements on single crystal samples, in particular along the lowest conductivity $a$-axis. This gives good reason to hope that polycrystalline GeSe samples with comparably low lattice thermal conductivity can be synthesized in the near future.

\subsection{Thermoelectric figure of merit in SnSe and GeSe}

The calculations of carrier transport can be combined with those of lattice thermal conductivity to make predictions for the thermoelectric figure of merit, $zT$.
The thermal conductivity due to hole transport, $\kappa_\mathrm{carr}$, was calculated using a first-principles framework \cite{chaves2021investigating,chaves2021microscopic} where the material dependent carrier concentration ($n_\mathrm{carr}$) and ionized impurity concentration ($n_\mathrm{ii})$ were determined in a self-consistent manner so that calculated values of the Seebeck coeffecient, $S$, and electrical conductivity, $\sigma$, matched experimental measurements in actual $p$-doped SnSe samples \cite{zhao2016intrinsic}. The temperature dependent carrier and impurity concentrations in SnSe were then used as a realistic approximation for the same quantities in GeSe, allowing for calculations of hole transport properties in GeSe under potential experimental conditions. 

\begin{figure}[tbp]
\includegraphics[width=0.49\columnwidth]{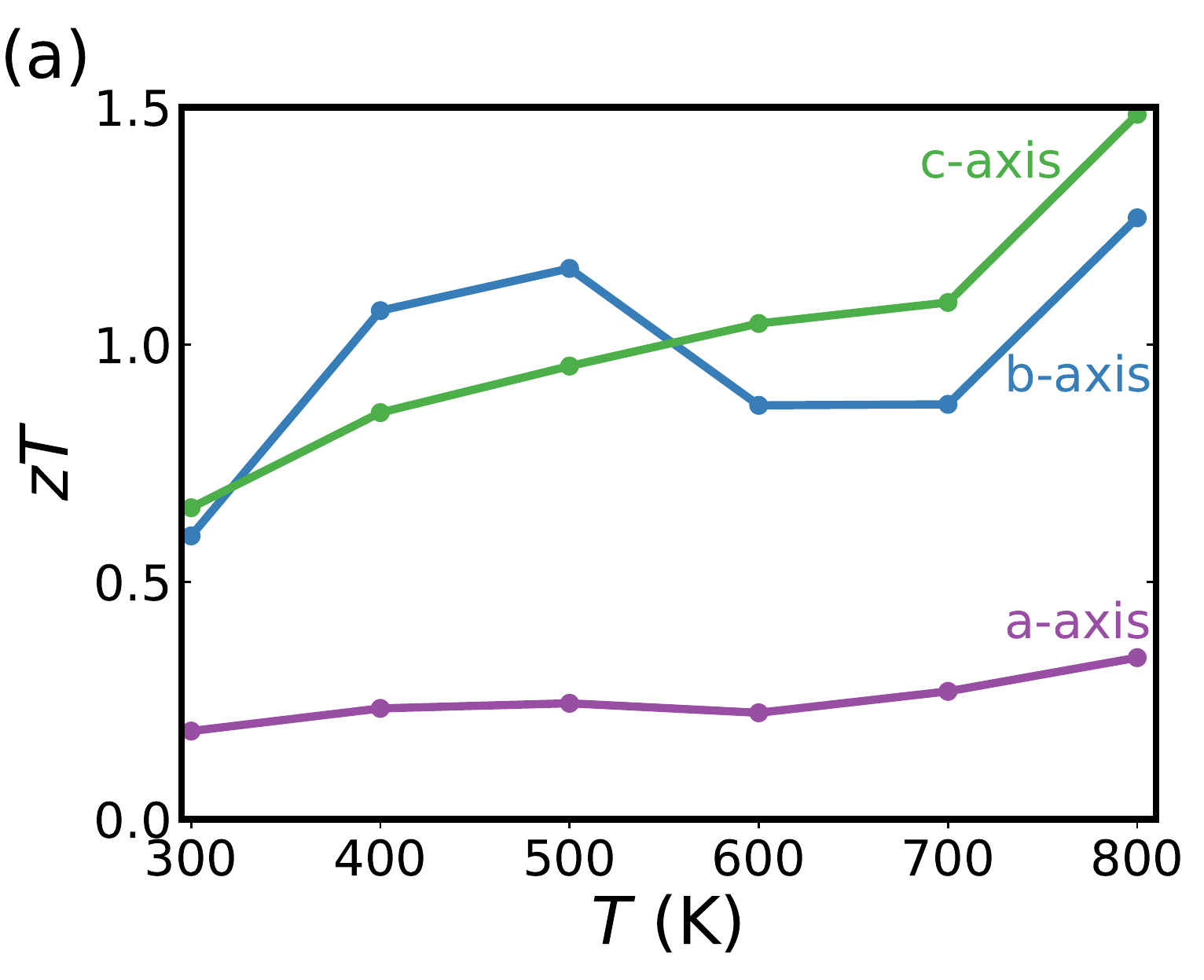}
\includegraphics[width=0.49\columnwidth]{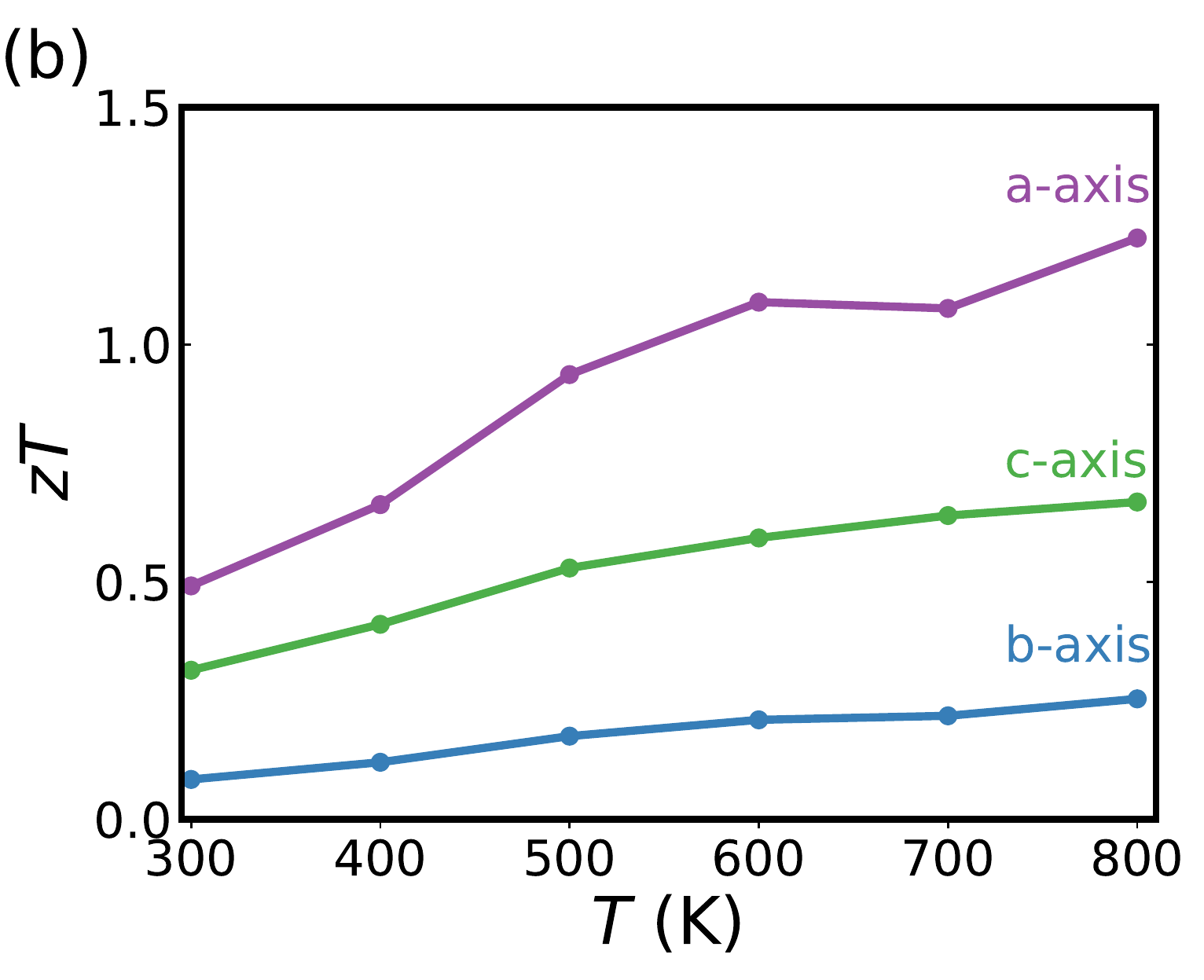}
\caption{\label{fig:GeSe-zT} Calculated thermoelectric figure of merit $zT = \sigma S^2 T/\kappa_\mathrm{tot}$ for each axis of hole-doped (a) SnSe and (b) GeSe, where all transport properties are calculated from first principles using carrier and impurity concentrations derived self-consistently from data for $a$-axis SnSe \cite{zhao2016intrinsic}.}
\end{figure}

The ultimate goal is discover practical thermoelectric materials, so in Fig.~\ref{fig:GeSe-zT} we combine the new calculations of lattice thermal conductivity with the carrier transport properties described in the previous paragraph, yielding a prediction for the thermoelectric figure of merit, $zT = \sigma S^2 T/\kappa_\mathrm{tot}$, along each of the axes of SnSe and GeSe.
The calculated values of $\kappa_\mathrm{tot}$ used here are higher than those based on the Debye-Callaway model, leading to less optimistic predictions for $zT$ than those presented in Ref.~\citenum{chaves2022out}. Nevertheless, GeSe exhibits significant thermoelectric potential, especially along the $a$-axis. This is in contrast to a previous prediction based on similar lattice thermal conductivity results but a simpler model for the carrier contributions, which yielded significantly lower $zT$ along the $a$-axis in comparison to the $c$-axis \cite{yuan2019tailoring}. Because our prediction is based on specific, realistic values of carrier and impurity concentrations, one can reasonably hope that by optimizing of the doping concentration the values of $zT$ can be improved\cite{chaves2022out}. Furthermore, polycrystalline samples of GeSe should, with proper care for purity and removal of oxides, yield equivalent or even lower thermal conductivities, and consequently even greater potential for a high figure of merit. 

\section{Conclusions}
\label{sec:conclusion}

We have outlined how, starting from the fundamental quantum mechanical Hamiltonian for a periodic crystal of nuclei and electrons, one can leverage modern high performance computing infrastructure to calculate a macroscopic property of significant technical importance such as the thermoelectric figure of merit, $zT$. 
The well-established framework of Kohn-Sham DFT converts the $N$-body Hamiltonian eigenvalue problem into a self-consistent numerical optimization problem for single-particle orbitals. 
This allows for the extraction of the electron-phonon and phonon-phonon couplings that determine the relevant scattering mechanisms that serve as input to the Boltzmann Transport Equation.
The electron and phonon distribution functions determined by the BTE can then be used to calculate all the necessary experimental observables that go into the calculation of $zT$.
As a demonstration of this process we focused on the thermoelectric performance of hole-doped SnSe and GeSe.
We have presented calculations of both the electronic transport properties (previously published in Refs.~\citenum{chaves2021microscopic,chaves2022out}) along with new calculations of the phonon lattice thermal conductivity and the resultant predictions for $zT$.

\section*{Statements and Declarations}

\textbf{Computational Resources} The calculations performed for this work used resources of CCJDR-IFGW-UNICAMP in Brazil, the National Energy Research Scientific Computing Center (NERSC), a U.S. Department of Energy Office of Science User Facility located at Lawrence Berkeley National Laboratory, operated under Contract No. DE-AC02-05CH11231, as well as the FASRC Cannon cluster supported by the FAS Division of Science Research Computing Group at Harvard University. 

\textbf{Funding} M.P.\ is supported by the Swiss National Science Foundation (SNSF) through the Early Postdoc.Mobility program (Grant No. P2ELP2-191706). A.A. gratefully acknowledges support from the Brazilian agencies CNPq and FAPESP under Grants No. 2010/16970-0, No. 2013/08293-7, No. 2015/26434-2, No. 2016/23891-6, No. 2017/26105-4, and No. 2019/26088-8. We acknowledge funding from the STC Center for Integrated Quantum Materials, NSF Grant No. DMR-1231319; NSF Award No. DMR-1922172; the Army Research Office under Cooperative Agreement Number W911NF-21-2-0147; and the Simons Foundation, Award No. 896626.

\textbf{Competing Interest} The authors have no relevant financial or non-financial interests to disclose.

\textbf{Author Contributions} All authors contributed to the study conception and design. Calculations and analysis of the data presented in the text were performed by Anderson S. Chaves and Daniel T. Larson. The first draft of the manuscript was written by Michele Pizzochero, Anderson S. Chaves, and Daniel T. Larson and all authors commented on previous versions of the manuscript. All authors read and approved the final manuscript.

\textbf{Availability of data}  The datasets generated during and/or analysed during the current study are available from the corresponding author on reasonable request.

\bibliography{references}
\end{document}